\documentclass[12pt]{article}

\usepackage{epsfig}
\usepackage{amssymb}

\newcommand{\beq}{\begin{equation}}
\newcommand{\bea}{\begin{eqnarray}}
\newcommand{\eeq}{\end{equation}}
\newcommand{\eea}{\end{eqnarray}}
\newcommand{\nn}{\nonumber\\}

\newcommand{\bx}{{\mathbf x}}

\newcommand\avr[1]{\left\langle{#1}\right\rangle}
\newcommand\avrc[1]{\left\langle{#1}\right\rangle_c}

\begin{document}

\title{\bf Lattice simulations of real-time quantum fields}

\author{
J.\ Berges$^1$, Sz.\ Bors\'anyi$^2$, D.\ Sexty$^1$,
I.-O.\ Stamatescu$^2$\\[0.5cm]
$^1$Institute for Nuclear Physics\\
Darmstadt University of Technology\\
Schlossgartenstr. 9, 64289 Darmstadt, Germany\\
$^2$Institute for Theoretical Physics\\
University of Heidelberg,\\
Philosophenweg 16, 69120 Heidelberg, Germany}

\date{}
\begin{titlepage}
\maketitle
\def\thepage{}          

\begin{abstract}
We investigate lattice simulations of scalar and nonabelian gauge fields 
in Minkowski space-time. For $SU(2)$ gauge-theory expectation values of link
variables in 3+1 dimensions are constructed by a stochastic
process in an additional (5th) ``Langevin-time''. A sufficiently
small Langevin step size and the use of a tilted real-time contour
leads to converging results in general. All fixed point solutions
are shown to fulfil the infinite hierarchy of
Dyson-Schwinger identities, however, they are not unique without
further constraints. For the nonabelian gauge theory the thermal
equilibrium fixed point is only approached at intermediate
Langevin-times. It becomes more stable if the complex time path is
deformed towards Euclidean space-time. We analyze this behavior
further using the real-time evolution of a quantum anharmonic
oscillator, which is alternatively solved by diagonalizing its
Hamiltonian. Without further optimization stochastic quantization 
can give accurate descriptions if the real-time extend of the lattice 
is small on the scale of the inverse temperature.
\end{abstract}

\end{titlepage}

\renewcommand{\thepage}{\arabic{page}}

\section{Introduction}

Lattice gauge theory provides important insight into strongly
interacting theories such as quantum chromodynamics (QCD).
Typically, calculations are based on a Euclidean formulation,
where the time variable is analytically continued to imaginary
values. By this the quantum theory is mapped onto a statistical
mechanics problem, which can be simulated by importance sampling
techniques. Recovering real-time properties from the Euclidean
formulation is a formidable problem that is still in its
infancies. Direct simulations in Minkowski space-time would be a
breakthrough in our efforts to resolve pressing questions, such as
early thermalization or the origin of seemingly perfect fluidity
in a QCD plasma at RHIC~\cite{HeinzRio}. For real times standard
importance sampling is not possible because of a non-positive
definite probability measure. Efforts to circumvent this problem
include mimicking the real-time dynamics by computer-time
evolution in Euclidean lattice
simulations~\cite{computertime,Miller:2000pd}. A problem in this
case is to calibrate the computer time independently of the
algorithm. Another procedure amounts
 to separate a positive factor from the Boltzmann factor to be used 
for importance sampling, and re-weight the configurations with the remaining
(complex) part. These so-called ``reweighting methods"
suffer from the problem of large cancellations of contributions, induced
by the oscillating signs in the weight (``sign problem") and from the  
difficulty of ensuring a sufficient overlap between the simulated and the 
target ensemble (``overlap problem"). 

Direct simulations in Minkowski space-time, however, may be
obtained using stochastic quantization techniques, which are not
based on a probability
interpretation~\cite{stochquant,Seiler:1983mz}. In Ref.~\cite{Berges:2005yt}
this has been recently used to explore nonequilibrium dynamics of
an interacting scalar quantum field theory.

In real-time stochastic quantization the quantum ensemble is
constructed by a stochastic process in an additional
``Langevin-time'' using the reformulation for the Minkowskian path
integral~\cite{cl,Minkowski,Huffel}: The quantum fields are
defined on a $d$-dimensional physical space-time lattice, and
the updating employs a Langevin equation with a complex
driving force in an additional, unphysical ``time'' direction.
This procedure does not
involve reweighting, nor redefinition of the Minkowski dynamics in
terms of an associated Euclidean one.
Though more or less formal proofs of equivalence of the stochastic
approach and the path integral formulation have been given for
Minkowski space-time, not much is known about the general
convergence properties and its reliability beyond free-field
theory or simple toy models~\cite{Minkowski,Huffel}. More advanced
applications concern simulations in Euclidean space-time with
non-real actions~\cite{Ambjorn:1986mf,Karsch:1985cb}. Besides
successful applications, major reported problems in this case
concern unstable Langevin dynamics and incidences of convergence
to "unphysical"
results~\cite{convergence,Ambjorn:1986mf,Karsch:1985cb}.

In this paper we discuss real-time stochastic quantization for
scalar field theory and $SU(N)$ pure gauge theory relevant for
QCD. Similar to what has been observed in
Ref.~\cite{Berges:2005yt} for scalar fields, also for the gauge
theory we find that previously reported unstable dynamics
represents no problem in practice. A combination of sufficiently
small Langevin step size and the use of a "tilted" real-time
contour leads to converging results in general. Our procedure
respects gauge invariance and appears to be well under control.
This is exemplified for $SU(2)$ gauge theory in $3+1$ dimensions.
For the scalar theory 
we consider, in particular, a zero-spatial dimension example, namely a
quantum anharmonic oscillator. The latter is solved in
addition by alternative methods using Hamiltonian diagonalization
for comparison. We find that stochastic quantization
can accurately describe the time evolution for lattices with
sufficiently small real-time extent. 
This concerns nonequilibrium
and equilibrium simulations at weak as well as strong couplings.
However, when the real-time extent of the lattice is enlarged the
stochastic updating does not converge to the correct solution.
In particular, for the nonabelian gauge theory the thermal
equilibrium solution is only approached at intermediate
Langevin-times. Its life-time in the course of the Langevin evolution
becomes longer if the complex time path is
deformed towards Euclidean space-time.

Since there appears more than one fixed point to which the
Langevin flow can converge, the solutions obtained by real-time
stochastic quantization are not unique in general. Similar
observations have been made before for Euclidean theories with
complex actions~\cite{convergence}. Euclidean theories with real actions can
be shown to have a unique solution based on positivity
arguments~\cite{stochquant}. A similar argument fails for real-time
stochastic quantization. In general, here the correct fixed point
cannot be chosen a priori without implementing further
constraints. We prove that all
possible fixed point solutions fulfil the same infinite set of
(symmetrized) Dyson-Schwinger identities of the quantum field
theory. However, solutions of Dyson-Schwinger equations without
specifying further constraints are not unique in general.
We discuss how tests for differentiating them can typically be 
devised and applied.

The paper is organized as follows. Sec.~\ref{sec:classical}
reviews stochastic quantization in Euclidean space-time, which
fixes the notation for the derivation of the real-time dynamics in
Sec.~\ref{sec:realdyn}. Particular emphasis is put on the
developments for gauge theories, since previous suggestions for
Langevin dynamics in the additional "time" dimension
violate Dyson-Schwinger identities of the original quantum theory.
In Sec.~\ref{sec:fixedpoints} we prove Dyson-Schwinger equations
to follow from real-time stochastic quantization. Simulation
results and precision tests are discussed in
Secs.~\ref{sec:prectest} and \ref{sec:testgauge}. We end with
conclusions in Sec.~\ref{sec:conclusions}.

\section{Euclidean stochastic quantization}
\label{sec:classical}

The stochastic quantization algorithm calculates ensemble averages
in an extended space of variables, where the lattice action $S$ of
the $d$-dimensional quantum theory plays the role of a potential
energy in a classical Hamiltonian $H$ of a ($d+1$)-dimensional, embedding
theory. The ensemble averages of the original quantum
theory are computed from classical evolution in the additional
"time" dimension.

\subsection{Euclidean scalar theory}
\label{sec:scalar}

We first consider classical dynamics for scalar fields in
$d$-dimensional Euclidean space-time with Hamiltonian~(see
e.g.~\cite{montvay}) \beq H =  \int\! {\rm d}^d x \left( \frac{1}{2}
\pi^2(x; t_5) + {\mathcal L_{\rm E}}(\varphi(x; t_5),\partial_x
\varphi(x; t_5)) \right)\, . \label{eq:hamiltonian} \eeq Here the
field $\varphi$ and momentum $\pi$ depend on the
\mbox{$d$-dimensional} Euclidean space-time variable $x_\mu$ ($\mu
= 0,\ldots 3$) and on an additional ``time'' variable~$ t_5$. The
function ${\mathcal L_{\rm E}}$ depends on the field and its
derivatives with respect to $x_\mu$. It is taken to correspond to
the Lagrange density of the physical four-dimensional Euclidean
field theory with mass $m$ and quartic self-interaction $\lambda$
where \beq {\mathcal L_{\rm E}} = \frac{1}{2} \left\{
\left(\partial_x \varphi(x; t_5)\right)^2 + m^2 \varphi^2(x; t_5)
\right\} + \frac{\lambda}{4!} \varphi^4(x; t_5) \,.
\label{eq:lagrangian} \eeq Since ${\mathcal L_{\rm}}$ contains no
field derivatives with respect to $ t_5$ the variable enters only
as an additional field label in Eq.~(\ref{eq:lagrangian}).

Expectation values of observables $F(\varphi)$ are given by the
functional integral
\begin{equation}
\langle F(\varphi) \rangle = Z^{-1} \int {\cal D}\pi{\cal
D}\varphi\, F(\varphi)\, e^{- H[\pi,\varphi]} \label{eq:expval}
\label{eq:canonical}
\end{equation}
with the normalization
\begin{equation} Z = \int {\cal D}\pi{\cal
D}\varphi\, e^{- H[\pi,\varphi]} \, .
\end{equation}
The Gaussian integrals over the momentum fields can be trivially
performed and the overall constants cancel out in the computation
of field expectation values. Here the introduction of the
canonical field momenta is used to compute expectation values of
quantum fields from classical Hamiltonian dynamics. This amounts
to replacing the canonical ensemble averages (\ref{eq:canonical})
by micro-canonical ones. The latter can be obtained from solving
Hamilton equations.

The basis for the numerical simulations are the discretized
classical equations for the Hamiltonian (\ref{eq:hamiltonian}).
The classical dynamics of the field $\varphi$ and its conjugate
momentum $\pi$ in $t_5$-time is described by Hamilton equations of
motion \bea \frac{\partial \varphi(x; t_5)}{\partial
 t_5} &=& \frac{\delta H}{\delta\pi(x; t_5)} \,\, = \,\,
\pi(x; t_5) \, , \label{eq:poissonq}
\\
\frac{\partial \pi(x; t_5)}{\partial  t_5} &=&
 - \frac{\delta H}{\delta\varphi(x; t_5)}
\,\, = \,\, - \frac{\delta S_{\rm E}[\varphi]}{\delta\varphi(x;
t_5)} \, , \label{eq:poissonp} \eea where the functional
differentiation with respect to $\varphi$ is for fixed $ t_5$. The
Euclidean action is \beq S_{\rm E}[\varphi]\, \equiv\, \int {\rm
d}^d x \, {\mathcal L_{\rm E}}(\varphi(x),\partial_x \varphi(x))
\, . \label{eq:euclidaction} \eeq

In order to discretize Eqs.~(\ref{eq:poissonq}) and (\ref{eq:poissonp})
in $t_5$, we denote the difference in subsequent $t_5$-time steps
by $\Delta  t_5 \equiv  (t_5)_{n+1} -  (t_5)_{n}$.
Correspondingly, we will write $\left\{\varphi( (t_5)_{n+1}),\pi(
(t_5)_{n+1})\right\} \to \left\{\varphi',\pi'\right\}$ and
$\left\{\varphi( (t_5)_{n}), \pi( (t_5)_{n})\right\} \to
\left\{\varphi,\pi \right\}$. A suitable second-order
discretization of (\ref{eq:poissonq}) then reads \bea \varphi'(x)
&=& \varphi(x) + \frac{\partial \varphi(x)}{\partial  t_5}\,
\Delta  t_5 + \frac{1}{2} \frac{\partial^2 \varphi(x)}{\partial
t_5^2}\, \Delta
 t_5^2
\nonumber\\[0.05cm]
&=& \varphi(x) + \pi(x) \, \Delta  t_5 - \frac{1}{2} \frac{\delta
S_E[\varphi]}{\delta \varphi(x)}\, \Delta  t_5^2 \, .
\label{eq:discretized} \eea Correspondingly, expanding $\pi'-\pi$
to order $\Delta  t_5$ and using Eq.~(\ref{eq:poissonp}) the field
momenta are (leapfrog) evolved with \beq \pi'(x) = \pi(x) -
\frac{1}{2} \left( \frac{\delta S_{\rm E}[\varphi]}{\delta
\varphi(x)} + \frac{\delta S_{\rm E}[\varphi']}{\delta
\varphi'(x)} \right)\, \Delta  t_5 \, , \label{eq:pis} \eeq which
yields reversible and area preserving discretized
dynamics~\cite{montvay}.

The conjugate momenta in the classical Hamiltonian
(\ref{eq:hamiltonian}) have a Gaussian distribution, which is
independent of the values of the field variables. If instead of
stepping along a single classical trajectory the momenta after
every single step are randomly refreshed, one recovers classical
Langevin dynamics in $ t_5$-time. It amounts to making the
substitutions \beq \sqrt{2}\, \pi(x) \,\to\, \eta(x)\, , \qquad
\frac{1}{2} \Delta  t_5^2 \,\to\, \epsilon \, ,
\label{eq:substitutions} \eeq with Gaussian noise \beq \langle
\eta(x) \rangle_\eta = 0 \,, \qquad \langle \eta(x)\, \eta(x')
\rangle_\eta = 2\, \delta(x-x') \, , \label{eq:noise} \eeq where
the average of an observable $A(\varphi;\eta)$ over the noise is
given by \beq \langle A(\varphi;\eta) \rangle_\eta \equiv
\frac{\int [{\rm d} \eta]\, A(\varphi;\eta)
\exp\left\{-\frac{1}{4} \int\! {\rm d}^4 x \, \eta^2(x)
\right\}}{\int [{\rm d} \eta]\, \exp\left\{-\frac{1}{4} \int\!
{\rm d}^4 x \, \eta^2(x) \right\}} \,\, . \label{eq:noiseeuklid}
\eeq The discretized Langevin equation then reads\footnote{Our
discretization corresponds to It\^{o} calculus for stochastic
processes.} \beq \varphi'(x) = \varphi(x) - \epsilon \,
\frac{\delta S_{\rm E}[\varphi]}{\delta \varphi(x)} +
\sqrt{\epsilon}\, \eta(x) \, . \label{eq:Langevin} \eeq The sum
over the Langevin steps, $\vartheta \equiv \sum \epsilon$, is
called "Langevin-time", which replaces $t_5$ of the classical dynamics.

This simplified description of the classical dynamics forms the
basis of stochastic quantization~\cite{stochquant}. The stochastic process
(\ref{eq:Langevin}) is associated to a distribution $P_{\rm E}(
t_5)$ for the field $\varphi(x)$. Writing $P_{\rm E}( (t_5)_{n+1})
\to P_{\rm E}^\prime$ and $P_{\rm E}( (t_5)_{n}) \to P_{\rm E}$
its evolution can be obtained from \bea P^\prime_{\rm
E}[\varphi^\prime] = \Bigg\langle \int [{\rm d}\varphi] P_{\rm
E}[\varphi]  \prod_{x} \delta\left(\varphi^\prime(x) - \varphi(x)
+ \epsilon \frac{\delta S_{\rm E}[\varphi]}{\delta \varphi(x)} -
\sqrt{\epsilon} \eta(x) \right) \Bigg\rangle_\eta  . \eea
Expanding the $\delta$-functionals and keeping only terms up to
order $\epsilon$ gives the Fokker-Planck equation \bea
\frac{1}{\epsilon} \left( P^\prime_{\rm E} - P_{\rm E}
\right)[\varphi] &=& \int\! {\rm d} t\, {\rm d}^3 x\,
\frac{\delta}{\delta \varphi} \left(\frac{\delta P_{\rm E}}{\delta
\varphi} + P_{\rm E}\, \frac{\delta S_{\rm E}}{\delta \varphi}
\right)[\varphi] + {\cal O}(\epsilon)\, , \label{eq:euklidfp} \eea
where we have used Eq.~(\ref{eq:noise}). For real action $S_{\rm E}$
one can prove that it converges to the late Langevin-time limit
\beq P_{\rm E}[\varphi] = e^{-S_{\rm E}[\varphi]} + {\cal
O}(\epsilon) \, . \label{eq:euclidsol} \eeq Therefore, expectation
values $\langle F(\varphi) \rangle$ can be obtained from noise
averages or, assuming ergodicity, from Langevin-time averages for
sufficiently long classical trajectories.

\subsection{Euclidean gauge theory}
\label{Sec:Enonabel}

A similar discussion can be done for gauge theories. We consider a
nonabelian pure gauge theory on an anisotropic lattice of size
$(N_s a_s)^3 \times N_\tau a_\tau$ with Euclidean action
\begin{eqnarray}
S_{\rm E}[U] &=& - \beta^0_{\rm E} \sum_{x} \sum_{i} \left\{
\frac{1}{2 {\rm Tr} {\bf 1}} \left( {\rm Tr}\, U_{x,0i} + {\rm
Tr}\, U_{x,0i}^{-1} \right) - 1 \right\}
\nonumber\\
&& - \beta^s_{\rm E} \sum_{x} \sum_{i,j \atop i<j} \left\{
\frac{1}{2 {\rm Tr} {\bf 1}} \left( {\rm Tr}\, U_{x,ij} + {\rm
Tr}\, U_{x,ij}^{-1} \right) - 1 \right\} \, ,
\label{eq:Splaquette}
\end{eqnarray}
with spatial indices $i,j = 1,2,3$. It is described in terms of
the gauge invariant plaquette variable
\begin{equation}
U_{x,\mu\nu} \equiv U_{x,\mu} U_{x+\hat\mu,\nu}
U^{-1}_{x+\hat\nu,\mu} U^{-1}_{x,\nu} \label{eq:plaq}\, ,
\end{equation}
where $U_{x,\nu\mu}^{-1}=U_{x,\mu\nu}\,$. Here $U_{x,\mu}$ is the
parallel transporter associated with the link from the neighboring
lattice point $x+\hat{\mu}$ to the point $x$ in the direction of
the lattice axis $\mu = 0,1,2,3$ with $U_{x,\mu} = U^{-1}_{x+
\hat\mu,-\mu}\,$. The link variable $U_{x,\mu}$ is an element of
the gauge group $G$. Because of the anisotropic lattice we have
introduced the anisotropic bare couplings $g_0$ for the time-like
plaquettes and $g_s$ for the space-like plaquettes with \beq
\beta^0_{\rm E} \equiv \frac{2 \gamma_{\rm E} {\rm Tr} {\bf
1}}{g_0^2} \,\, , \quad \beta^s_{\rm E} \equiv \frac{2 {\rm Tr}
{\bf 1}}{g_s^2 \gamma_{\rm E}} \, , \label{eq:ganisoE} \eeq where
$\gamma_{\rm E} \equiv a_s/a_\tau$ is the anisotropy parameter.

For $G = SU(N)$ one has $U_{x,\mu}^{-1} = U^\dagger_{x,\mu}$. When
we consider Minkowski space-time below we will observe that the
latter no longer holds. Therefore, we keep $U^{-1}_{x,\mu\nu}$ in
the definition of the action (\ref{eq:Splaquette}), which will
still be valid for the Minkowskian theory.

In order to derive the $t_5$-time dynamics of the 5-dimensional
theory one can follow the equivalent steps as in
Sec.~\ref{sec:scalar}. For the definition of the Langevin drift
term one has to define differentiation with respect to the
nonabelian variable $U_{x,\mu}$. Differentiation in group space
will be defined by \beq D_{x \mu a} f(U_{x,\mu}) =
\frac{\partial}{\partial \alpha} f\left( e^{i \alpha \lambda_a}
U_{x,\mu}\right)|_{\alpha = 0} \eeq with the generators
$\lambda_a$ of the Lie algebra. For $G = SU(N)$ these are the
$a=1,\ldots,N^2-1$ traceless, hermitian $N \times N$ Gell-Mann
matrices, which are normalized to ${\rm Tr}(\lambda_a \lambda_b) =
2 \delta_{ab}$ with $[\lambda_a,\lambda_b] = 2 i f_{abc}
\lambda_c$, where the structure constants $f_{abc}$ are completely
antisymmetric and real. The derivatives then satisfy the
commutation relations $[D_{x \mu a}, D_{x \mu b}] = 2 f_{abc} D_{x
\mu c}$.

For the Hamiltonian \beq H = {\frac{1}{2}} \sum_{x \mu a} P_{x \mu
a}^2( t_5) + S_{\rm E}[U( t_5)] \eeq with the above definitions
the Hamilton equations for the $ t_5$-time dynamics read \beq
\frac{\partial U_{x, \mu}}{\partial  t_5} = i \lambda_a P_{x \mu
a} U_{x, \mu} \quad , \quad \frac{\partial P_{x \mu a}}{\partial
 t_5} = - D_{x \mu a} S_{\rm E}[U] \label{eq:derivU}\, . \eeq
The discretized canonical equations of motion corresponding to
(\ref{eq:discretized}) and (\ref{eq:pis}) are then given by \bea
U^{\prime}_{x, \mu} &=& U_{x, \mu} + \frac{\partial U_{x,
\mu}}{\partial  t_5} \Delta  t_5 + \frac{1}{2} \frac{\partial^2
U_{x, \mu}}{\partial  t_5^2} \, \Delta  t_5^2
\nonumber\\
&=& U_{x, \mu} + i \lambda_a P_{x \mu a} U_{x, \mu}\, \Delta  t_5
+ \frac{1}{2} \frac{\partial}{\partial  t_5} \left(i \lambda_a
P_{x \mu a} U_{x, \mu} \right)\, \Delta  t_5^2
\nonumber\\
&=& \exp\left\{ i \lambda_a \left( P_{x \mu a} \Delta  t_5 -
\frac{1}{2} D_{x \mu a} S_{\rm E}[U]\, \Delta  t_5^2
\right)\right\} U_{x, \mu}\, , \label{eq:A} \eea where the
equalities hold up to corrections ${\cal O}(\Delta t_5^3)$. Here
$P_{x \mu}^a$ is the $ t_5$-time conjugate momentum for
$U_{x,\mu}$ with (leapfrog) discretized Hamilton equation
(cf.~(\ref{eq:pis})) \bea P^{\prime}_{x \mu a} &=& P_{x \mu a} -
\frac{1}{2} \left( D_{x \mu a} S_{\rm E}[U] + D'_{x \mu a} S_{\rm
E}[U'] \right)\, \Delta  t_5 \, , \label{eq:P} \eea where $D'$
refers to the derivative with respect to $U'$.

The Langevin equation is obtained in the same way as described for
the scalar case by the corresponding substitutions
(\ref{eq:substitutions}) with
\begin{equation}
\sqrt{2}\, P_{x \mu a}(x) \,\to\, \eta_{x \mu a}(x) \, ,
\label{eq:replaceP}
\end{equation}
and a real Gaussian noise $\eta_{x \mu a}$ satisfying \beq \langle
\eta_{x \mu a} \rangle_\eta = 0 \,, \qquad \langle \eta_{x \mu
a}\, \eta_{y \nu b} \rangle_\eta = 2\, \delta_{\mu\nu} \delta_{xy}
\delta_{ab} \, . \label{eq:Pnoise} \eeq The discretized Langevin
equations for $U_{x, \mu}$ to this order in $\epsilon$ may then be
written as \bea U^{\prime}_{x, \mu} &=& \exp\left\{ - i \lambda_a
\left(\epsilon D_{x \mu a} S_{\rm E}[U] - \sqrt{\epsilon}\,
\eta_{x \mu a} \right)\right\} U_{x, \mu}\, .
\label{eq:ALangevinE} \eea

\section{Real-time dynamics}
\label{sec:realdyn}

\subsection{Real-time scalar theory}
\label{sec:realscalar}

Instead of the embedded $d$-dimensional Euclidean theory discussed
above, one may consider Minkowskian space-time. This requires
an analytic continuation of the Euclidean time in the action
(\ref{eq:euclidaction}) to Minkowski time. For this it is
instructive to consider the one-parameter family of actions
$S_\xi$ for the field $\varphi(x)$ with $x \equiv (t,\bx)$ given
by \beq S_\xi[\varphi] = - \int\! {\rm d}^d x\,  e^{-i \pi \xi/2}
\left\{\frac{1}{2} \varphi \square_\xi \varphi + \frac{1}{2} m^2
\varphi^2 + \frac{\lambda}{4!} \varphi^4 \right\} \, ,
\label{eq:minact} \eeq with d'Alembertian \beq \square_\xi \equiv
e^{i \pi \xi} \partial_t^2 - \vec\nabla^2 \, . \eeq For $\xi = 0$
one recovers from Eq.~(\ref{eq:minact}) the standard Minkowski action,
while for $\xi = 1$ one has \beq S_{\xi = 1} \,\equiv\, i\, S_{\rm
E} \, . \eeq

The discretized Langevin equation (\ref{eq:Langevin}) can be
written for the family of actions (\ref{eq:minact}) as
\begin{equation}
\varphi'(x) = \varphi(x) + i\, \epsilon \, \frac{\delta
S_\xi[\varphi]}{\delta \varphi(x)} + \sqrt{\epsilon}\, \eta(x) \,
. \label{eq:complexlange}
\end{equation}
with \beq \frac{\delta
S_\xi[\varphi]}{\delta \varphi(x)} = e^{-i \pi \xi/2} \left\{
\square_\xi \varphi(x) + m^2 \varphi(x) + \frac{\lambda}{3!}
\varphi^3(x) \right\}\,.\eeq
The Gaussian noise $\eta(x)$ is defined as in the Euclidean case in
Eq.~(\ref{eq:noise}).
The case $\xi = 0$
forms the basis of real-time stochastic quantization~\cite{Minkowski}, and
we will denote $S_{\xi = 0} \equiv S$.

It is important to note that possible solutions of
Eq.~(\ref{eq:complexlange}) will not be real in general. For instance,
for a real scalar field theory with Minkowskian action
$S[\varphi]$ it will generate complex field values for $ t_5
> 0$. For a complex field \beq \varphi(t,\bx) = \varphi_R(t,\bx) +
i \varphi_I(t,\bx) \eeq also the conjugate momenta $\pi(t,\bx)$
or, with (\ref{eq:substitutions}), the respective noise \beq
\eta(t,\bx) = \eta_R(t,\bx) + i \eta_I(t,\bx) \eeq can be complex.
Equation (\ref{eq:complexlange}) may then be written as \bea
\varphi^\prime_R(t,\bx) &=& \varphi_R(t,\bx) - \epsilon \,
I_\xi(\varphi_R,\varphi_I;t,\bx) + \sqrt{\epsilon}\, \eta_R(t,\bx)
\, ,
\nonumber\\
\varphi^\prime_I(t,\bx) &=& \varphi_I(t,\bx) + \epsilon \,
R_\xi(\varphi_R,\varphi_I;t,\bx) + \sqrt{\epsilon}\, \eta_I(t,\bx)
\, , \label{eq:langecomponents} \eea where \bea
R_\xi(\varphi_R,\varphi_I;t,\bx) &\equiv& {\rm
Re}\left(\frac{\delta S_\xi[\varphi]}{\delta \varphi(t,\bx)}
\Big|_{\varphi =\varphi_R + i \varphi_I} \right) \, , \,\,
\nonumber\\
I_\xi(\varphi_R,\varphi_I;t,\bx) &\equiv& {\rm
Im}\left(\frac{\delta S_\xi[\varphi]}{\delta \varphi(t,\bx)}
\Big|_{\varphi =\varphi_R + i \varphi_I}\right) \, . \label{eq:RI}
\eea It remains to consider suitable choices of the noise terms. A
particularly simple choice is a real noise, where $\eta_I \equiv
0$. However, for complex fields a real noise may not be suitable
in general, since the noise plays the role of the field conjugate
momentum in the underlying classical dynamics (see
(\ref{eq:substitutions})).
For a complex noise one obtains from
(\ref{eq:noise})\footnote{Note that the complex field $\varphi$
yields a non-real action $S[\varphi]$, which is not a functional
of $\varphi^*\varphi$. Likewise, the conjugate momentum or complex
noise does not fulfil $\langle \eta^*(x)\, \eta(x') \rangle_\eta
= 2 \, \delta(x-x')$ but (\ref{eq:noise}). } \bea &&\!\!\!\!\!\!
\langle \eta_R(x) \rangle_\eta \,=\, 0 \,, \quad \langle \eta_I(x)
\rangle_\eta \,=\, 0 \, , \quad \langle \eta_R(x)\, \eta_I(x')
\rangle_\eta \,=\, 0 \, ,
\nonumber\\
&&\!\!\!\!\!\! \langle \eta_R(x)\, \eta_R(x') \rangle_\eta \,=\, 2
\alpha  \delta(x-x') \, , \quad \langle \eta_I(x)\, \eta_I(x')
\rangle_\eta \,=\, 2 (\alpha - 1) \delta(x-x') \, . \qquad
\label{eq:comnoise} \eea For real $\alpha > 1$ the average of an
observable $A(\varphi;\eta)$ over the noise is defined as \beq
\langle A \rangle_\eta \equiv \frac{\int [{\rm d} \eta_R] [{\rm d}
\eta_I]\, A \exp\left\{- \int\! {\rm d}^4 x \, \left( \frac{1}{4
\alpha}\, \eta_R^2 + \frac{1}{4(\alpha-1)}\,
\eta_I^2\right)\right\}} {\int [{\rm d} \eta_R] [{\rm d} \eta_I]\,
\exp\left\{- \int\! {\rm d}^4 x \, \left(\frac{1}{4 \alpha}\,
\eta_R^2 + \frac{1}{4(\alpha-1)}\, \eta_I^2 \right) \right\}} \, .
\eeq The case $\alpha = 1$ corresponds to a real noise with
$\eta_I \equiv 0$ and noise average as in (\ref{eq:noiseeuklid})
with $\eta \to \eta_R$. We will see in Sec.~\ref{sec:fixedpoints}
that the different possible choices do not affect the final
result.

Other observables $F(\varphi(x))$ follow similar evolution
equations, which are obtained like (\ref{eq:discretized}) from
discretization to second order in $\Delta  t_5$,  \beq
F\left(\varphi'(x)\right) = F\left(\varphi(x)\right) +
\frac{\partial F\left(\varphi(x)\right)}{\partial  t_5}\, \Delta
 t_5 + \frac{1}{2} \frac{\partial^2
F\left(\varphi(x)\right)}{\partial  t_5^2}\, \Delta  t_5^2
\label{eq:scalarF}\, ,\eeq with the substitutions
(\ref{eq:substitutions}) and using (\ref{eq:complexlange}). E.g.\
the product of two fields follows the discretized equation \bea
\varphi'(x)\varphi'(y) &=& \varphi(x)\varphi(y) + i \epsilon \,
\left\{\frac{\delta S_\xi[\varphi]}{\delta \varphi(x)}\varphi(y) +
\frac{\delta S_\xi[\varphi]}{\delta \varphi(y)}\varphi(x) \right\}
\nonumber\\
&+& \epsilon \, \eta(x)\eta(y) +
\sqrt{\epsilon}\,\left\{\varphi(x) \eta(y) + \varphi(y)
\eta(x)\right\} . \quad \label{eq:Langevin2pt} \eea For three
powers of the field one finds \bea \varphi'(x)\varphi'(y)
\varphi'(z) &=& \varphi(x)\varphi(y) \varphi(z)
\nonumber\\
&+& \sqrt{\epsilon} \left\{ \varphi(x)\varphi(y) \eta(z)
+\varphi(x) \varphi(z) \eta(y)+ \varphi(y)\varphi(z)\eta(x)
\right\}
\nonumber\\
&-& \epsilon \left\{ \frac{\delta S_\xi[\varphi]}{\delta
\varphi(x)}\varphi(y)\varphi(z) +\frac{\delta
S_\xi[\varphi]}{\delta \varphi(y)}\varphi(x)\varphi(z)
+\frac{\delta S_\xi[\varphi]}{\delta
\varphi(z)}\varphi(x)\varphi(y) \right\}
\nonumber\\
&+&\epsilon \left\{ \varphi(x)\eta(y)\eta(z) +
\varphi(y)\eta(x)\eta(z) + \varphi(z)\eta(x)\eta(y) \right\}\, ,
\label{eq:threepoint} \eea and correspondingly for higher powers
of the field. In Sec.~\ref{sec:fixedpoints} we will see that the
late Langevin-time limit of these equations leads to an infinite
hierarchy of Dyson-Schwinger identities.

\subsection{Real-time gauge theory}
\label{sec:nonabelianrealtime}

Following Sec.~\ref{sec:realscalar} we replace the Euclidean
action (\ref{eq:Splaquette}) by the Minkowskian with
\begin{equation}
 -S_{\rm E}[U]= i S[U]  \label{eq:minkS}
\end{equation}
on a real-time lattice of size $(N_s a_s)^3 \times N_t a_t$. The
real-time classical action reads
\begin{eqnarray}
S[U] &=& - \beta_0 \sum_{x} \sum_i \left\{ \frac{1}{2 {\rm Tr}
{\bf 1}} \left( {\rm Tr}\, U_{x,0i} + {\rm Tr}\, U_{x,0i}^{-1}
\right) - 1 \right\}
\nonumber\\
&& + \beta_s \sum_{x} \sum_{i,j \atop i<j} \left\{ \frac{1}{2 {\rm
Tr} {\bf 1}} \left( {\rm Tr}\, U_{x,ij} + {\rm Tr}\, U_{x,ij}^{-1}
\right) - 1 \right\} \, ,
\label{eq:clgaugeaction}
\end{eqnarray}
where the relative sign between the time-like and the space-like
plaquette terms reflects the Minkowski metric, and
\begin{equation}
\beta_0 \equiv \frac{2 \gamma {\rm Tr} {\bf 1}}{g_0^2} \,\, ,
\quad \beta_s \equiv \frac{2 {\rm Tr} {\bf 1}}{g_s^2 \gamma} \, ,
\label{eq:ganisoM}
\end{equation}
with the anisotropy parameter $\gamma \equiv
a_s/a_t$.\footnote{See Sect~\ref{Sec:Enonabel} for $a_t = -i
a_\tau$ and the replacement according to Eq.~(\ref{eq:minkS}).}

The discretized Langevin equations for $U_{x, \mu}$ to second order in
$\epsilon$ may then be written as
\begin{eqnarray}
U^{\prime}_{x, \mu} &=& \exp\left\{ i \lambda_a \left(\epsilon\, i
D_{x \mu a} S[U] + \sqrt{\epsilon}\, \eta_{x \mu a}
\right)\right\} U_{x, \mu}\, , \label{eq:ALangevinM}
\end{eqnarray}
with
\begin{eqnarray}
i D_{x \mu a} S[U] &=&  \frac{1}{2N} \beta_0
\sum_{j} \left\{ \delta_{j\mu} {\rm Tr} \left( \lambda_a U_{x,j}
C_{x,j0}
-\bar{C}_{x,j0} U^{-1}_{x,j} \lambda_a \right) \right.
\nonumber\\
&& \left. + \delta_{0\mu} {\rm Tr} \left( \lambda_a U_{x,0}
C_{x,0j} - \bar{C}_{x,0j}
U^{-1}_{x,0} \lambda_a  \right) \right\}
\nonumber\\
&& - \frac{1}{2N} \beta_s \sum_{i,j \atop i \neq j} \delta_{j\mu}
{\rm Tr} \left( \lambda_aU_{x,j} C_{x,ji}- \bar{C}_{x,ji}
U^{-1}_{x,j} \lambda_a \right)
\nonumber\\
&=& - \frac{1}{2N} \sum_{\nu=0 \atop \nu \neq \mu}^{3}
\beta_{\mu\nu} {\rm Tr} \left( \lambda_a U_{x,\mu} C_{x,\mu\nu}-
\bar{C}_{x,\mu\nu} U^{-1}_{x,\mu} \lambda_a \right) \, .
\label{eq:derivS}
\end{eqnarray}
For a compact notation we have defined $\beta_{ij} \equiv \beta_s$
and $\beta_{0i} \equiv \beta_{i0} \equiv - \beta_0$ and
\begin{eqnarray}
C_{x,\mu\nu} &=& U_{x+\hat\mu,\nu} U^{-1}_{x+\hat\nu,\mu}
U^{-1}_{x,\nu} + U^{-1}_{x+\hat\mu-\hat\nu,\nu}
U^{-1}_{x-\hat\nu,\mu} U_{x-\hat\nu,\nu}
\nonumber\\
\bar{C}_{x,\mu\nu} &=&
U_{x,\nu}U_{x+\hat\nu,\mu}U^{-1}_{x+\hat\mu,\nu} +
U^{-1}_{x-\hat\nu,\nu} U_{x-\hat\nu,\mu}U_{x+\hat\mu-\hat\nu,\nu}
\, .
\end{eqnarray}
With $U_{x,\mu} C_{x,\mu\nu} =  U_{x,\mu\nu} + U_{x,\mu(-\nu)}$
and $\bar{C}_{x,\mu\nu} U^{-1}_{x,\mu} = U_{x,\mu\nu}^{-1} +
U_{x,\mu (-\nu)}^{-1}$ one observes that the sum in
Eq.~(\ref{eq:derivS}) is over all possible plaquettes containing
$U_{x,\mu}\,$.

Specifying to $SU(2)$ gauge theory $\lambda_a=\sigma_a$
($a=1,2,3$) represent the Pauli matrices, and we can make further
simplifications using $ {\rm Tr} (U^{-1} \sigma^a) = -{\rm Tr} ( U
\sigma^a)$ for any element $U \in SU(2)$. The latter
simplification also holds for $U \in SL(2,{\bf C})$. This is
relevant since, similar to what has been discussed for the scalar
theory above, possible solutions of Eq.~(\ref{eq:ALangevinM}) may
respect an enlarged symmetry group. Taking \beq U_{x,\mu} \equiv
e^{i A_{x \mu a} \sigma_a/2} \eeq the vector fields $A_{x \mu a}$
need not to be real, which is in contrast to the Euclidean case
discussed in Sec.~\ref{Sec:Enonabel}. The complex matrix $A_{x
\mu}^a \sigma_a$ still remains traceless, however, the Hermiticity
properties are lost. As a consequence, it is no longer possible to
identify $U^{\dagger}$ with $U^{-1}$ as is taken into account in
Eq.~(\ref{eq:Splaquette}). This corresponds to an extension of the
original $SU(2)$ manifold to $SL(2,{\bf C})$ for the Langevin
dynamics. Only after taking noise averages the original $SU(2)$
gauge theory is to be recovered (see Sec.~\ref{sec:fixedpoints}).

According to Eq.~(\ref{eq:Pnoise}) the noise is given by
\begin{equation}
\langle \eta_{x \mu a} \rangle_\eta = 0 \,, \qquad \langle \eta_{x
\mu a}\, \eta_{y \nu b} \rangle_\eta = 2\, \delta_{\mu\nu}
\delta_{xy} \delta_{ab} \, . \label{eq:realtimenoise}
\end{equation}
It is essential to use the same statistics for the noise as one
has in the Euclidean case. One may be tempted to replace the
$\delta_{\mu\nu}$ on the right-hand side of
Eq.~(\ref{eq:realtimenoise}) by $g_{\mu\nu}$ to make the Langevin
equation manifestly covariant~\cite{Huffel}. However, in this case
solutions of the Langevin evolution would not respect the
Dyson-Schwinger identities of the underlying quantum field theory,
as is shown in section \ref{sec:fixedpoints}. Only for observables
one has to respect Lorentz symmetries which is still fulfilled
with Eq.~(\ref{eq:realtimenoise}).

In general, observables $F(U)$ follow similar evolution equations,
which are obtained from the second-order discretization equivalent
to Eq.~(\ref{eq:scalarF}). For instance, for the plaquette
$U_{x,\mu\nu}$ given by (\ref{eq:plaq}) one has \beq
U_{x,\mu\nu}^\prime = U_{x,\mu\nu} + \frac{\partial
U_{x,\mu\nu}}{\partial  t_5}\, \Delta  t_5 + \frac{1}{2}
\frac{\partial^2 U_{x,\mu\nu}}{\partial  t_5^2}\, \Delta
 t_5^2 \label{eq:langplaq}\, ,\eeq with Eqs.~(\ref{eq:derivU}) and
(\ref{eq:minkS}). This will be used to derive the corresponding
Dyson-Schwinger equation for the plaquette variable in
Sec.~\ref{sec:DSU}.

\section{Fixed points of the Langevin flow in real time}
\label{sec:fixedpoints}

\subsection{Scalar theory}

Starting from some (arbitrary) initial field value $\varphi(x)$ at
$ t_5 = 0$ we compute the Langevin-time flow according to equation
(\ref{eq:complexlange}). A fixed point of the evolution equation
is defined by the condition for the noise-averaged field $\langle
\varphi'(x) \rangle_\eta - \langle \varphi(x) \rangle_\eta = 0$.
It simultaneously corresponds to a fixed point in the space of all
noise-averaged correlation functions
$\langle\varphi(x)\varphi(y)\rangle_\eta$,
$\langle\varphi(x)\varphi(y)\varphi(z)\rangle_\eta$, etc.\
following equations (\ref{eq:Langevin2pt}), (\ref{eq:threepoint})
and similarly for higher $n$-point functions. According to
Eq.~(\ref{eq:complexlange}) a fixed point for the one-point function
corresponds to \bea \left\langle \frac{\delta
S_\xi[\varphi]}{\delta \varphi(x)}\right\rangle_\eta &=& 0 \,,
\label{eq:sd1} \eea where we have used Eq.~(\ref{eq:noise}) or
equivalently Eq.~(\ref{eq:comnoise}). It is important to note that the
result is independent of the different possible implementations of
the noise according to Eq.~(\ref{eq:comnoise}). The two-point function
described by Eq.~(\ref{eq:Langevin2pt}) then fulfils
\begin{equation}
\left\langle \frac{\delta S_\xi[\varphi]}{\delta
\varphi(x)}\varphi(y) \right\rangle_\eta + \left\langle
\frac{\delta S_\xi[\varphi]}{\delta \varphi(y)}\varphi(x)
\right\rangle_\eta = 2 i \delta(x-y) \, . \label{eq:sd2}
\end{equation}
For the 3-point function one has from
(\ref{eq:threepoint}) \bea \left\langle\frac{\delta
S_\xi[\varphi]}{\delta \varphi(x)}\varphi(y)\varphi(z)
\right\rangle_\eta +\left\langle \frac{\delta
S_\xi[\varphi]}{\delta \varphi(y)}\varphi(x)\varphi(z)
\right\rangle_\eta +\left\langle \frac{\delta
S_\xi[\varphi]}{\delta
\varphi(z)}\varphi(x)\varphi(y) \right\rangle_\eta =&& \nonumber\\
2 i \left(\left\langle \varphi(x) \right\rangle_\eta \delta(y-z) +
\left\langle \varphi(y) \right\rangle_\eta \delta(x-z) +
\left\langle \varphi(z) \right\rangle_\eta \delta(x-y)\right) \, ,
\label{eq:sd3} \eea and correspondingly for the higher $n$-point
functions. Equations (\ref{eq:sd1})--(\ref{eq:sd3}) and the
corresponding equations for $n > 3$ are the symmetrized
Dyson-Schwinger identities for the time-ordered correlation
functions of the scalar theory in Minkowski space-time.

\subsection{Gauge theory}
\label{sec:DSU}

For the plaquette variable (\ref{eq:langplaq}) one has
\begin{eqnarray}
\left\langle \frac{\partial^2 U_{x,\mu\nu}}{\partial  t_5^2}
\right\rangle_\eta &=& \left\langle \frac{\partial^2
U_{x,\mu}}{\partial  t_5^2} U_{x+\hat\mu,\nu}
U^{-1}_{x+\hat\nu,\mu} U^{-1}_{x,\nu} + U_{x,\mu} \frac{\partial^2
U_{x+\hat\mu,\nu}}{\partial  t_5^2}
U^{-1}_{x+\hat\nu,\mu} U^{-1}_{x,\nu} \right. \nonumber\\
&+& \left. U_{x,\mu} U_{x+\hat\mu,\nu} \frac{\partial^2
U^{-1}_{x+\hat\nu,\mu}}{\partial  t_5^2} U^{-1}_{x,\nu} +
U_{x,\mu} U_{x+\hat\mu,\nu} U^{-1}_{x+\hat\nu,\mu}
\frac{\partial^2 U^{-1}_{x,\nu}}{\partial  t_5^2}
\right\rangle_\eta \nonumber\\
&\stackrel{!}{=}&0 \, , \label{eq:DSgauge1}
\end{eqnarray}
where we have used that noise averages over products of
first-order \mbox{$ t_5$-derivatives} vanish. The latter can be
observed from the fact that link and momentum variables are not
correlated such that their noise average factorizes. According to
Eq.~(\ref{eq:realtimenoise}) the averages over the corresponding
momenta vanish identically. The symmetrized Dyson-Schwinger
equation for the plaquette variable then follows from
Eqs.~(\ref{eq:derivU}), (\ref{eq:minkS}) and (\ref{eq:replaceP}), i.e.\
\begin{eqnarray}
\frac{\partial^2 U_{x,\mu}}{\partial  t_5^2} &=& i \sum_{a}
\lambda_a \left(i D_{x \mu a} S[U]\right) U_{x,\mu}-
\frac{1}{2}\sum_{a,b}
\lambda_a \eta_{x \mu a}\lambda_b \eta_{x \mu b} U_{x,\mu} \, ,\nonumber\\
\frac{\partial^2 U^{-1}_{x,\mu}}{\partial  t_5^2} &=& -i
U^{-1}_{x,\mu} \sum_{a} \left(i D_{x \mu a} S[U]\right) \lambda_a
- \frac{1}{2} U^{-1}_{x,\mu} \sum_{a,b} \lambda_a \eta_{x \mu
a}\lambda_b \eta_{x \mu b} \, ,
\end{eqnarray}
with Eq.~(\ref{eq:realtimenoise}) and
\begin{equation}
\sum_{a} \left(\lambda_a\right)_{\alpha\beta}
\left(\lambda_a\right)_{\gamma\delta} = 2
\left(\delta_{\alpha\delta} \delta_{\beta\gamma} -
\frac{1}{N}\delta_{\alpha\beta} \delta_{\gamma\delta}\right) \, .
\end{equation}
By setting the first term of the sum in Eq.~(\ref{eq:DSgauge1}) to
zero and taking the trace, one obtains, using the notation of
Eq.~(\ref{eq:derivS}):
\begin{eqnarray}
2 \frac{N^2-1}{N} \langle {\rm Tr} U_{x,\mu\nu}
\rangle_\eta =
 \frac{i}{N} \sum_{\gamma=0 \atop \gamma \neq \mu}^{3}
 \beta_{\mu\gamma}
\Bigg\langle  {\rm Tr} \left( \bar{C}_{x,\mu\gamma} U^{-1}_{x,\mu}
U_{x,\mu\nu} -
U_{x,\mu} C_{x,\mu\gamma} U_{x,\mu\nu} \right) &&
\nonumber\\
 - \frac{1}{N} {\rm Tr} U_{x,\mu\nu} {\rm Tr} \left( \bar{C}_{x,\mu\gamma}
 U^{-1}_{x,\mu} - U_{x,\mu}
C_{x,\mu\gamma}\right) \Bigg\rangle_\eta  &&
 \, . \label{eq:DSpre}
\end{eqnarray}

\begin{figure}[t]
\begin{center}
\epsfig{file=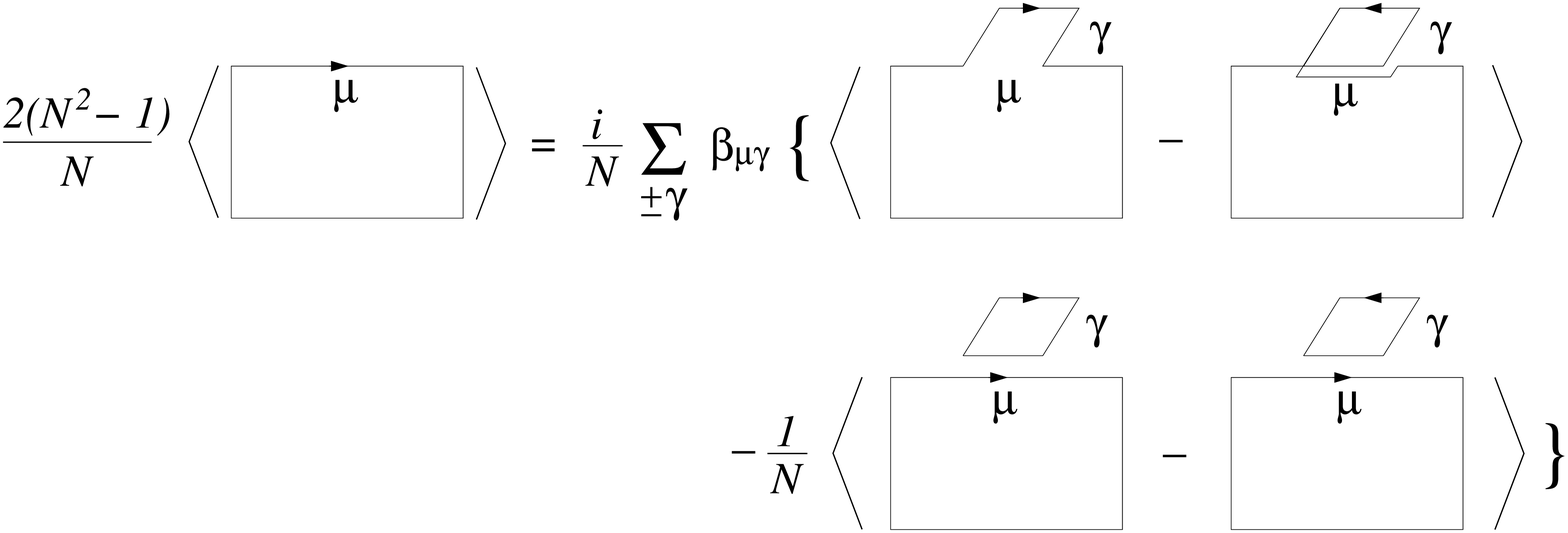,width=12.cm}
\end{center}
\caption{Graphical representation of the Dyson-Schwinger equation
for a Wilson loop.} \label{fig:DS}
\end{figure}

The first term on the RHS of this equation contains $U_{x,\mu}$
and its inverse such that the loop can be viewed as including a
departure in the $\gamma$-direction.  The second term contains
$U_{x,\mu}$ twice: The loop can be viewed as including a "crossed
path" (see Fig.~\ref{fig:DS}).  The last terms contain separately
traced plaquettes. For $U \in SL(2,{\bf C})$ or $SU(2)$ one has
${\rm Tr} U = {\rm Tr} U^{-1}$ such that these last terms in
Eq.~(\ref{eq:DSpre}) vanishes, i.e.\ ${\rm Tr} \left( U_{x,\mu}
C_{x,\mu\nu} - \bar{C}_{x,\mu\nu} U^{-1}_{x,\mu} \right) = 0$ in
this case. A similar analysis can also be done for the second to
fourth terms of Eq.~(\ref{eq:DSgauge1}), such that the sum is
symmetric in all the links of the plaquette $U_{x,\mu\nu}\,$. In
Fig.~\ref{fig:DS} the Dyson-Schwinger equation is displayed
graphically, where it is generalized to an arbitrary closed loop.
When a certain link appears more than once in the original loop
then additional "contact" terms appear. The respective analysis
follows along the same lines and will not be considered here.

Fixed points of the Langevin flow for observables $F[U]$
representing gauge-invariant products of link variables along
closed loops are obtained from $\langle
\partial^2 F[U]/\partial t_5^2 \rangle_\eta = 0$. The corresponding
set of Dyson-Schwinger identities \cite{xue} constitute an infinite
system of equations whose solution is not unique in general
without specifying further boundary conditions. Accordingly, we
will typically find more than one possible fixed point below, when
we solve the Langevin equation numerically. They all solve the
same set of Dyson-Schwinger equations. In contrast to the case of
Euclidean space-time, where the associated Fokker-Planck equation
(\ref{eq:euklidfp}) can be shown to converge to a unique solution
at late Langevin-time, there seems no general proof for real
times, i.e.\ in the absence of a positive definite probability
weight.

\section{Accuracy tests I: scalar theory}
\label{sec:prectest}

\begin{figure}[t]
\begin{center}
\epsfig{file=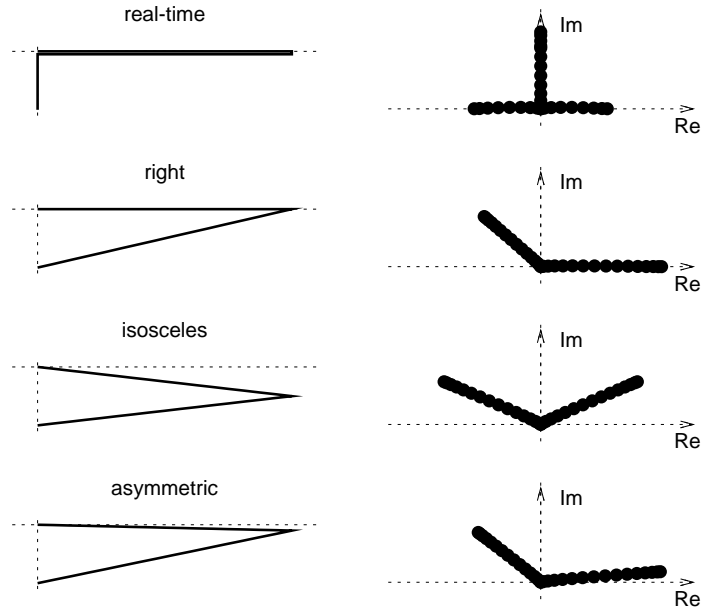,width=10.cm}
\end{center}
\caption{Complex time-contours (left) and the corresponding
distribution of eigenvalues (right) characterizing the Langevin
dynamics for a free scalar theory as described in the main text.
}
\label{fig:contourev}
\end{figure}

In thermal equilibrium the fields obey the periodicity condition
$\varphi(0,\bx) = \varphi(-i\beta,\bx)$ with inverse temperature
$\beta$. Accordingly, correlation functions in Euclidean
space-time can be computed using a purely imaginary time-path from
$t=0$ to $-i\beta$. Thermal equilibrium correlation functions
$\langle \varphi(x_1)\varphi(x_2)\ldots\varphi(x_n)\rangle$ with
real times $t_1, t_2, \ldots t_n$ have to be computed using a
time-path that extends along the real-time axis including these
times. The curves on the left of Fig.~\ref{fig:contourev} give
some examples of possible real-time contours for thermal
equilibrium along with other complex contours that will be
employed below. The curves (also denoted as "closed" and "rectangular")
both first run along the real-time axis and then turn in different
ways to $-i\beta$. The curves "isosceles" and "asymmetric"
exhibit a tilt with respect to the real-time axis. For
nonequilibrium evolutions we will use the isosceles triangle or
asymmetric time-path below, where a small tilt with respect to the
real-time axis can serve as a regulator to improve convergence.

We consider field theory for the different choices of complex
contours shown in Fig.~\ref{fig:contourev}. We discretize the
contours using the complex-time points $C_t$ with $t=0,\ldots,N_t$
and ${\rm Re} C_0 = {\rm Re} C_{N_t}$. With the notation
$\varphi_t(\bx) \equiv \varphi(C_t,\bx)$ and $\Delta_t \equiv
C_{t+1}-C_t$ a scalar theory on such a contour is defined by the
action
\begin{eqnarray}
S &=& \frac{1}{2} \sum_{t}\int d^3x \Bigg\{
\frac{(\varphi_{t+1}(\bx)-\varphi_t(\bx))^2}{\Delta_t}
\nonumber\\
&& + \frac{\Delta_t}{2} \left[\varphi_{t+1}(\bx) \vec\nabla^2
\varphi_{t+1}(\bx)+\varphi_{t}(\bx)
\vec\nabla^2 \varphi_{t}(\bx) \right] \nonumber\\
&& - \Delta_t \left[ V\left(\varphi_{t+1}(x)\right) +
V\left(\varphi_t(x)\right) \right] \Bigg\} \, .
\label{eq:complexS}
\end{eqnarray}
It is instructive to consider for a moment the free theory with
$V(\varphi)= m^2\varphi^2/2$ and neglect spatial dimensions for
simplicity. For the free theory the action (\ref{eq:complexS}) is
quadratic in the fields:
\begin{equation}
S=\frac{1}{2}\sum_{t,t'}\varphi_t G^{-1}_{tt'} \varphi_t' \quad ,
\qquad \frac{\partial^2 S}{\partial\varphi_t \partial\varphi_{t'}}
= G^{-1}_{tt'} \, . \label{eq:prop}
\end{equation}
The complex inverse contour propagator $G^{-1}_{tt'}$ is symmetric
and not Hermitian in general. With
\begin{equation}
\sum_{t'}G^{-1}_{tt'}\psi^{a}_{t'}=c^a \psi^{a}_{t}
\end{equation}
its eigenvectors $\psi^{a}_t$ and complex eigenvalues $c^a$ can be
used to write the action as
\begin{equation}
S = \frac{1}{2}\sum_a c^a \chi^a \chi^a \quad, \qquad \chi^a =
\sum_t \psi^a_t \varphi_t \, ,
\end{equation}
where we have used completeness and orthogonality relations
$\sum_t\psi^{a}_t\psi^{b}_t = \delta^{ab}$ and
$\sum_a\psi^{a}_t\psi^{a}_{t'} = \delta_{tt'}$.

In Fig.~\ref{fig:contourev} we show the distribution of complex
eigenvalues $c^a$ for the different corresponding contours
displayed on the left. The eigenvalues can become small though
they are non-zero. They depend strongly on the chosen contour. Its
impact on the Langevin dynamics can be observed from equation
(\ref{eq:complexlange}), which becomes here a set of independent
Langevin equations
\begin{equation}
\chi^{a\prime} = \chi^a + i \epsilon c^a \chi^a +\sqrt{\epsilon}
\eta^a \quad , \qquad \langle\eta^a\eta^b\rangle=2\delta^{ab} \, ,
\label{eq:reducedL}
\end{equation}
with $\chi^a=\sum_t \psi^a_t \eta_t$. For a purely imaginary
time-contour from zero to $-i\beta$ one has ${\rm Re} c^a=0$,
${\rm Im} c^a > 0$ ($\forall a$), as for the Euclidean case described by
Eq.~(\ref{eq:Langevin}).
The Langevin evolution in $t_5$-time converges for ${\rm Im} c^a >
0$ \cite{Minkowski} and one finds in the continuum limit
\begin{equation}
\langle \chi^a \chi^b\rangle_\eta + \frac{2 \delta^{ab}}{i
(c^a+c^b)} \sim e^{i(c^a+c^b) \vartheta} \, .
\end{equation}
Therefore, eigenvalues with small positive imaginary part can
converge slowly. However, this linear or free field theory
analysis becomes invalid in the presence of sufficiently strong
interactions. The associated damping due to interactions in
realistic field theories typically leads to rapid convergence, as
can be observed from the numerical results below. (See also
Ref.~\cite{Berges:2005yt}.)

Fluctuations can grow large if the dynamics is governed
by small real and imaginary part of the eigenvalues $c^a$. The
linear analysis indicates that this should be a problem for
contours along the real-time axis. From the eigenvalue
distribution of Fig.~\ref{fig:contourev} one observes that a
non-vanishing tilt of the contour with respect to the real-time
axis may serve as a regulator. For our simulations we employ a
small tilt such that the contour always proceeds downwards, i.e.\
${\rm Im} C_t > {\rm Im} C_{t+1} \forall t$.
 Without or for very small regulator
we encounter incidences of unstable Langevin dynamics (see also
\cite{convergence}). Their appearance depends on the random
numbers and they are strongly suppressed by using a smaller
Langevin step size. We normally discard these trajectories and
typically employ a Langevin step size of $\epsilon \sim 10^{-5}$.
\begin{figure}[t]
\begin{center}
\epsfig{file=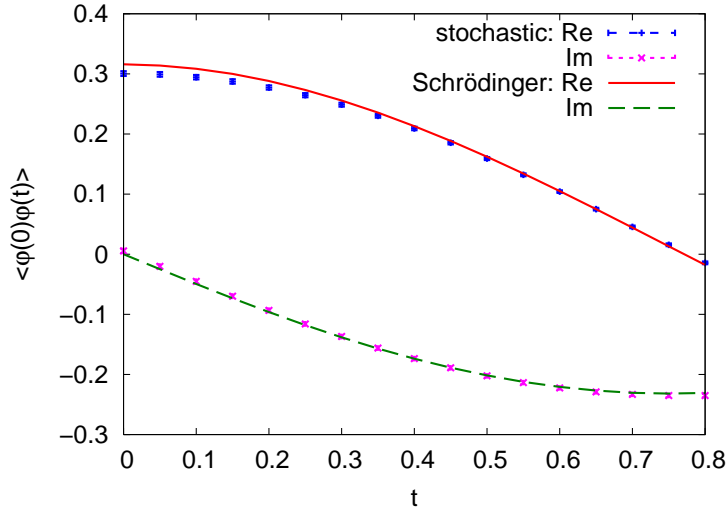,width=10.cm}
\end{center}
\caption{Real and imaginary part of the unequal-time two-point
correlation function for the quantum anharmonic oscillator as a
function of real time $t$. The results obtained from stochastic
quantization agree to very good accuracy to those obtained from
directly solving the Schr{\"o}dinger equation.} \label{fig:T4dh}
\end{figure}

\subsection{Short-time evolution: Thermal fixed point}
\label{sec:thermal}

We consider an interacting scalar theory with classical action
(\ref{eq:complexS}), where
\begin{equation}
V(\varphi)= \frac{1}{2} m^2\varphi^2 + \frac{\lambda}{4!}
\varphi^4 \, , \label{eq:effpot}
\end{equation}
in zero spatial dimension. The Schr{\"o}dinger equation for the
corresponding quantum anharmonic oscillator can be solved
numerically by diagonalization of the Hamiltonian. We use the
occupation number representation and truncate
the Hilbert space keeping 32 dimensions corresponding to the
lowest occupation numbers. Exponentiation of the diagonalized
Hamiltonian yields the time translation operator, which may be
used for real as well as complex times. We checked that the
results are insensitive to an enlargement of the truncation
dimension. This is compared to the results obtained from
stochastic quantization by solving the Langevin dynamics according
to (\ref{eq:complexlange}). In the following all quantities will
be given in appropriate units of the mass parameter $m$. We first
perform simulations in thermal equilibrium, in which case the time
contour extends along the imaginary axis to the inverse
temperature $\beta$ (see Fig.~\ref{fig:contourev}). Non-equilibrium
is considered in Sec.~\ref{sec:noneq}.
\begin{figure}[t]
\begin{center}
\epsfig{file=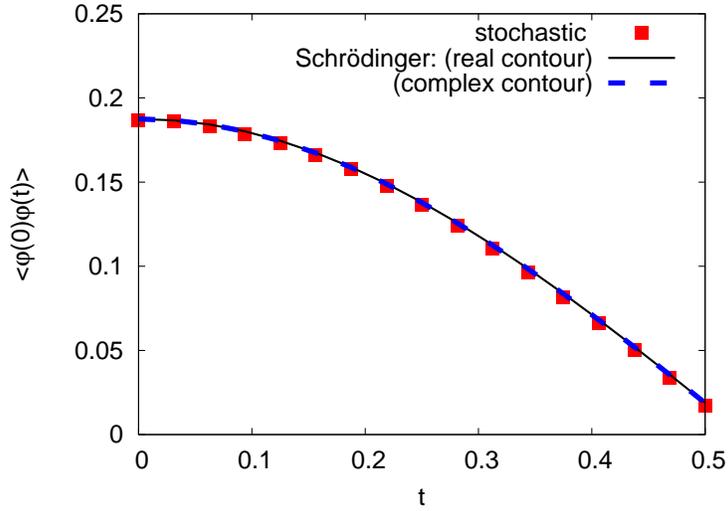,width=10.cm}
\end{center}
\caption{Similar evolution as in Figs.~\ref{fig:T4dh} but for much
stronger coupling $\lambda = 96$ and $\beta = 1$.} \label{fig:T4i}
\end{figure}

\begin{figure}[t]
\begin{center}
\epsfig{file=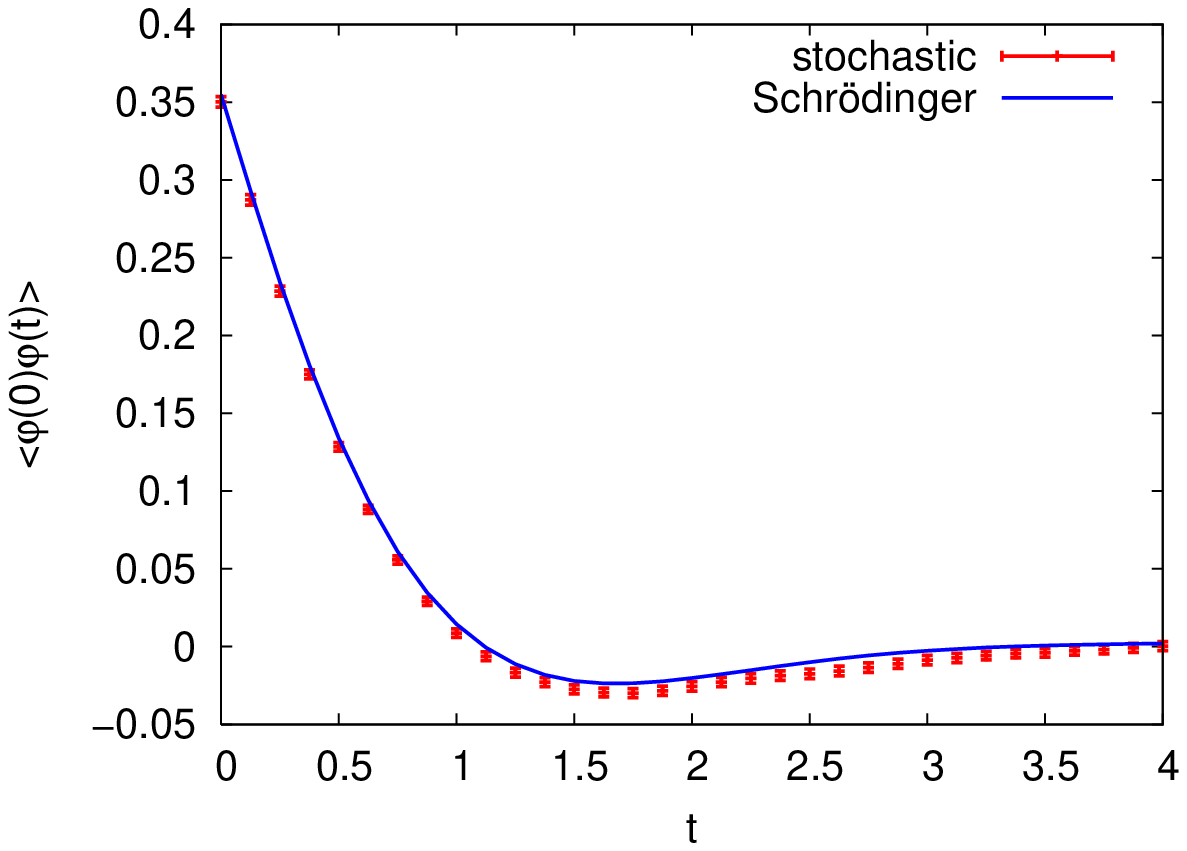,width=10.cm}
\end{center}
\caption{Similar evolution as in Figs.~\ref{fig:T4dh} but for
smaller temperature ($\beta = 8$ and $\lambda = 6$) on an isosceles
triangle contour.} \label{fig:T5k}
\end{figure}

Fig.~\ref{fig:T4dh} shows the two-point correlation function
$\langle \varphi(0) \varphi(t) \rangle_\eta$ as a function of real
time $t$ for $\lambda = 24$ and $\beta = 1$. The real-time extent
of the contour is $t_{\rm final} = 0.8$, such that the upper branch has
a tilt of $0.01 \beta$ and the lower branch has a tilt of $0.99
\beta$. Square symbols denote the results from stochastic
quantization for the real part of the two-pint function and
crosses for the imaginary part, respectively. For comparison the
corresponding results from the Schr{\"o}dinger equation are given,
which agree well.\footnote{For Figs.~\ref{fig:T4dh}--\ref{fig:T5k}
the total number of points along the contour ranges from $32$ to
$64$, with equal number of points on each branch. The
noise-average is obtained from the average over $10^3$ to about
$10^4$ runs. The Langevin step size is
$\epsilon \sim 10^{-4}$ to $10^{-5}$.} 

Fig.~\ref{fig:T4i} shows the real part of the two-point
correlation function as a function of real time $t$, where we
employ a much stronger coupling $\lambda = 96$ and $\beta = 1$.
Square symbols denote the results from stochastic quantization.
Here the real-time extent of the time contour is $t_{\rm final} = 0.5$.
The upper branch of the contour has a tilt of $0.001 \beta$ with
respect to the real-time axis, so it is almost horizontal and,
therefore, realizes to high accuracy a real-time contour. For
comparison results from the Schr{\"o}dinger equation are displayed
as well. The solid line corresponds to Minkowski time evolution
("real contour"), whereas the dashed line gives the results for
those complex times with small imaginary part as employed in the
stochastic quantization simulations ("complex contour"). They all
agree well. We fit the time evolution to
\begin{equation}
\langle \varphi(0) \varphi(t) \rangle = a \cos \left(\omega \,
{\rm Re} t\right) e^{-\gamma {\rm Re} t} \, , \label{eq:fitd}
\end{equation}
which allows us to extract characteristic damping times. We find
from the stochastic quantization simulation $\gamma/\omega =
0.018$ and from the Schr{\"o}dinger equation $\gamma/\omega =
0.013$ (Minkowski) and $\gamma/\omega = 0.014$ on the contour. The
frequency is always $\omega = 2.95$.

Fig.~\ref{fig:T5k} shows a time evolution for smaller temperature
with $\beta = 8$ and $\lambda = 6$. The real-time extent of the
contour is $t_{\rm final} = 4$. Using the fit (\ref{eq:fitd}) we find
from the stochastic quantization simulation $\gamma/\omega = 0.92$
and from the Hamiltonian diagonalization method $\gamma/\omega =
1.0$ on the contour. The frequency is $\omega = 1.4$.

\begin{figure}[t]
\begin{center}
\epsfig{file=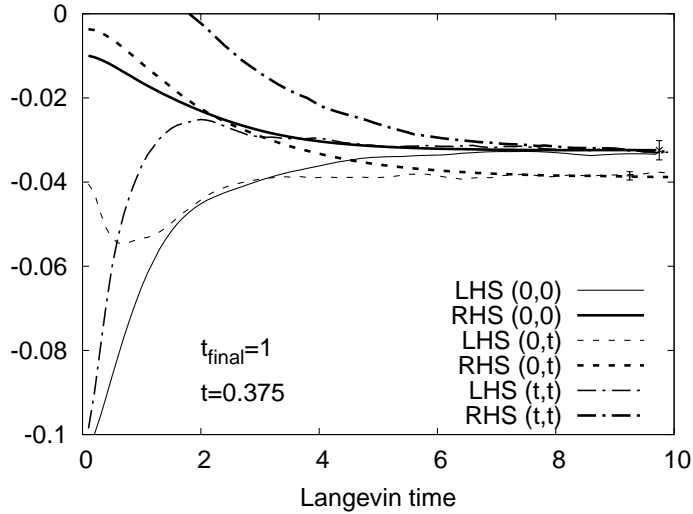,width=10.cm}
\end{center}
\caption{Shown is the Langevin-time evolution of the two-point
function and four-point function given by the LHS and RHS of the
Dyson-Schwinger identity (\ref{eq:DS2}). The curves are for fixed
real-time values. At sufficiently late Langevin-time they agree to
very good accuracy, thus becoming related by the Dyson-Schwinger
equation. } \label{fig:sd_TS1d}
\end{figure}

Fig.~\ref{fig:sd_TS1d} verifies the validity of the
Dyson-Schwinger identity (\ref{eq:sd2}) for various real-time
values. We use the parameters $\lambda=24$ and $\beta=1$. The
real-time extent of the contour is $t_{\rm final}=1$ with 32
points in total on the (isosceles triangle) contour. Plotted is the
Langevin-time evolution of the LHS and the RHS of the
equation\footnote{Eq.\ (\ref{eq:sd2}) is the symmetrized form of
(\ref{eq:DS2}).}
\begin{equation}
\sum_{\bar{t}} G^{-1}_{0,t \bar{t}}\langle
\varphi_{\bar{t}}\varphi_{t'} \rangle - \delta_{tt'} = -i
\frac{\lambda}{3!} \langle \varphi_{t} \varphi_{t} \varphi_{t}
\varphi_{t'} \rangle \, ,\label{eq:DS2}
\end{equation}
where $G^{-1}_{0,t \bar{t}}$ is the inverse propagator
(\ref{eq:prop}) of the free theory.  One observes that at
sufficiently late Langevin times both sides agree to very good
accuracy. In particular, one sees that equal-time values at
different real times agree, which has to be the case for the time
translation invariant thermal solution.

\subsection{Long-time evolution: Non-unitary fixed points}

\begin{figure}[t]
\begin{center}
\epsfig{file=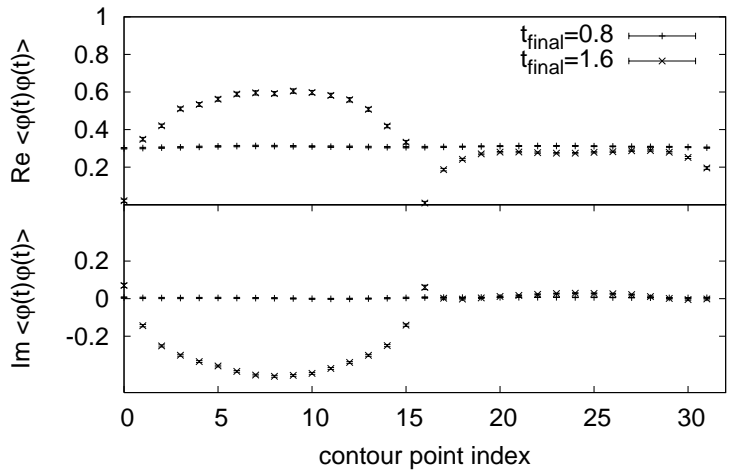,width=10.cm}
\end{center}
\caption{Real and imaginary part of the equal-time correlator as a
function of time (index of lattice site along the time-contour)
for the parameters ($\lambda = 24$ and $\beta = 1$) as well as the contour of
Fig.~\ref{fig:T4dh}. Compared are two simulations where the
real-time extent of the lattice differs by a factor of two. The
larger lattice leads to a qualitatively different, non-unitary
behavior.} \label{fig:mean_T4dhp}
\end{figure}

We observed above that stochastic quantization can describe the
real-time evolution very accurately for short times. However, when
the real-time extent of the lattice is enlarged the stochastic
updating may not converge to the correct solution. As an example,
Fig.~\ref{fig:mean_T4dhp} shows the two-point correlation function
as a function of real time $t$ for $\lambda = 24$ and $\beta = 1$,
i.e.\ for the parameters of Fig.~\ref{fig:T4dh}. From that figure
we have seen that for a real-time lattice with $t_{\rm final}=
0.8$ there is excellent agreement of stochastic quantization and
Schr{\"o}dinger equation results. Correspondingly,
Fig.~\ref{fig:mean_T4dhp} exhibits a constant real part and a
vanishing imaginary part of the equal-time correlator in thermal
equilibrium for $t_{\rm final}= 0.8$. However, doubling the extent
of the contour leads to a qualitatively different behavior as
indicated by the cross symbols of Fig.~\ref{fig:T4dh}. This
difference persists also on finer real-time grids. The
non-vanishing imaginary part of the equal-time correlator and the
loss of time-translation invariance reflects a non-unitary time
evolution.

\begin{figure}[t]
\begin{center}
\epsfig{file=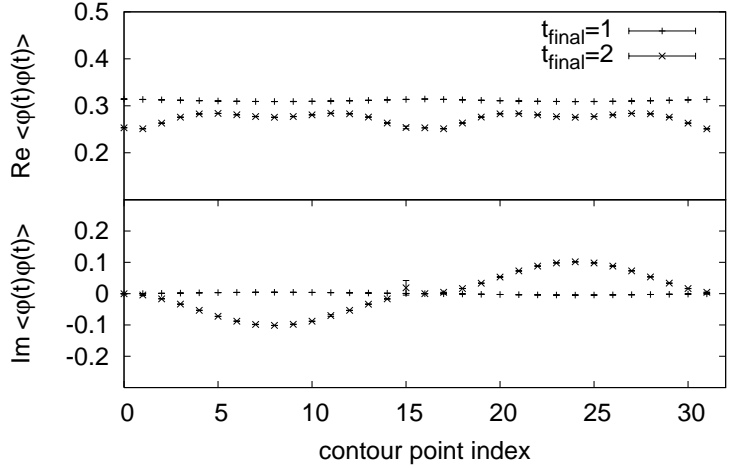,width=10.cm}
\end{center}
\caption{Similar evolution as in Fig.~\ref{fig:mean_T4dhp},
but on an isosceles triangle contour. While
the thermal solution obtained for $t_{\rm final} = 1$ does not
depend on the contour details (see also Fig.~\ref{fig:sd_TS1d}),
the properties of the non-unitary
fixed point obtained for $t_{\rm final} = 2$ are contour
dependent (see also Figs.~\ref{fig:sd2s_TS2c} and \ref{fig:sd2_TS2c}).} \label{fig:mean_TS1d2c}
\end{figure}

The properties of these non-unitary fixed point solutions depend
on the details of the time contour. This is in contrast to the
universal properties of the thermal solution. As an example,
Fig.~\ref{fig:mean_TS1d2c} shows the equal-time two point function
for same $\lambda=24$ and $\beta=1$ as in
Fig.~\ref{fig:mean_T4dhp}, however, with different contour geometry
(isosceles triangle). For $t_{\rm final} = 1$ one still observes
the thermal fixed point solution, whose properties agree very well
with those obtained from lattices with smaller $t_{\rm final}$.
Stretching the temporal extent of the contour to $t_{\rm final}=2$
the Langevin dynamics converges to a non-unitary fixed point as
one can see from Fig.~\ref{fig:mean_TS1d2c}. The detailed
properties of this fixed point differ from those shown in
Fig.~\ref{fig:mean_T4dhp}.

\begin{figure}[t]
\begin{center}
\epsfig{file=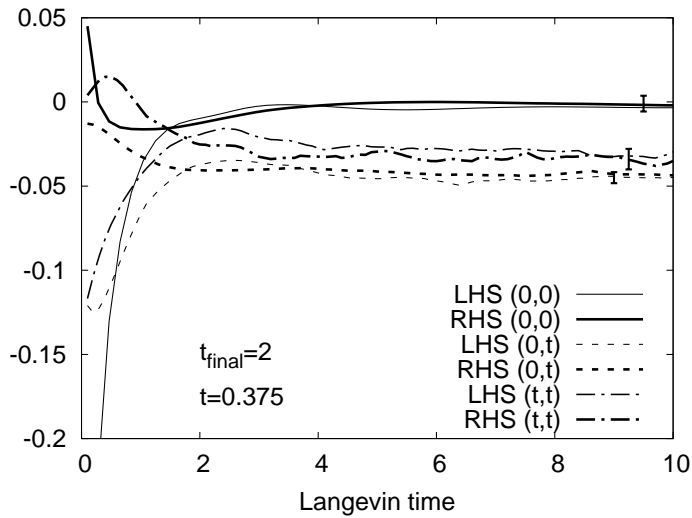,width=10.cm}
\end{center}
\caption{The left-hand-side (LHS) and right-hand-side (RHS) of the
Dyson-Schwinger equation (\ref{eq:DS2symm}) for the non-unitary
fixed point, which agree well at late Langevin time. Contrary to
the properties of the thermal fixed point (see
Fig.~\ref{fig:sd_TS1d}), the $(0,0)$ and $(t,t)$ components are
not equal, which reflects the loss of time translation
invariance.} \label{fig:sd2s_TS2c}
\end{figure}

\begin{figure}[t]
\begin{center}
\epsfig{file=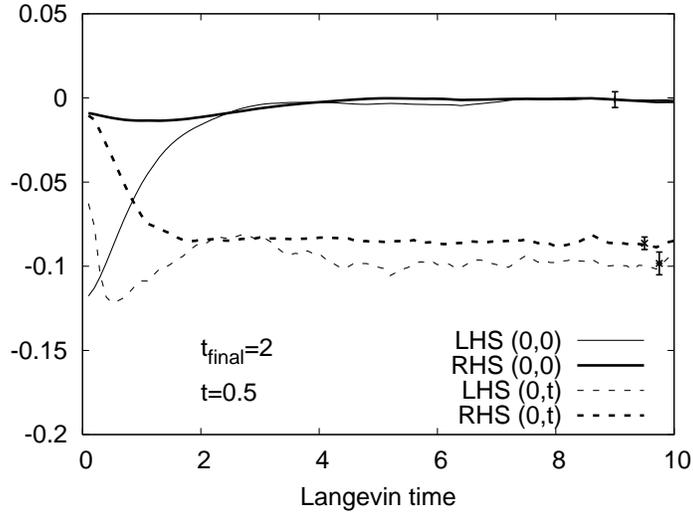,width=10.cm}
\end{center}
\caption{ The left-hand-side (LHS) and right-hand-side (RHS) of
equation (\ref{eq:DS2}), showing a discrepancy beyond the
statistical error.} \label{fig:sd2_TS2c}
\end{figure}

In Fig.~\ref{fig:sd_TS1d} we have shown how the propagator
fulfils for the thermal fixed point the corresponding
Dyson-Schwinger identity. According to the results of
Sec.~\ref{sec:fixedpoints} any fixed point solution has to fulfil
the same properly symmetrized Dyson-Schwinger identities. For
instance, the identity~(\ref{eq:sd2}) reads
\begin{equation}
\sum_{\bar{t}} G^{-1}_{0,t \bar{t}}\langle
\varphi_{\bar{t}}\varphi_{t'} \rangle + \sum_{\bar{t}}
G^{-1}_{0,t' \bar{t}} \langle \varphi_{\bar{t}}\varphi_{t} \rangle
+ -2\delta_{tt'} = - i\frac{\lambda}{3!} \langle \varphi_{t}
\varphi_{t} \varphi_{t} \varphi_{t'} \rangle - i\frac{\lambda}{3!}
\langle \varphi_{t} \varphi_{t'} \varphi_{t'} \varphi_{t'} \rangle
\, .\label{eq:DS2symm}
\end{equation}
We display the LHS and the RHS of this equation as a function of
Langevin-time in Fig.~\ref{fig:sd2s_TS2c}. Indeed, one observes
that the symmetrized Dyson-Schwinger equation (\ref{eq:sd2}) holds
within statistical errors at late Langevin-time for the
non-unitary solution. In Fig.~\ref{fig:sd2_TS2c} we check
Eq.~(\ref{eq:DS2}), which is the non-symmetrized version of
Eq.~(\ref{eq:sd2}). Here the discrepancy we find for the
non-unitary fixed point is greater than the statistical error. In
contrast, the thermal fixed point solution respects, of course,
both Eqs.~(\ref{eq:DS2symm}) and (\ref{eq:DS2}) as required by any
physical solution for the underlying quantum field theory.

\subsection{Non-equilibrium dynamics}
\label{sec:noneq}

Non-equilibrium dynamics can be described by the generating
functional for correlation functions~\cite{Berges:2004yj}:
\begin{eqnarray}
\lefteqn{Z[J;\rho] =
{\rm Tr}\left\{ \rho\, T_{\mathcal C}\, e^{i \int_{\mathcal C}\!
J(x) \Phi(x)}\right\}} \nonumber\\
&=& \int {\rm d} \varphi_1 {\rm d} \varphi_2 \,
\rho(\varphi_1,\varphi_2) \int\limits_{\varphi_1}^{\varphi_2}
D^{\prime} \varphi\, e^{i \int_{\mathcal C} \left( L(x) +
J(x)\varphi(x)\right)} \label{eq:definingZneq}  .\,\,
\end{eqnarray}
The path integral (\ref{eq:definingZneq}) displays the quantum
fluctuations for a theory with Lagrangian $L$, and the statistical
fluctuations encoded in the weighted average with the initial-time
(non-thermal) density matrix $\rho(\varphi_1,\varphi_2)$. Here
$T_{\mathcal C}$ denotes contour time ordering along a closed path
$\mathcal C$ starting at $t \equiv x^0 =0$ with
\mbox{$\int_{\mathcal C} \equiv \int_{\mathcal C} {\rm d} x^0 \int
{\rm d}^d x$}. This corresponds to usual time ordering along the
forward piece $\mathcal C_+$, and anti-temporal ordering on the
backward piece $\mathcal C_-$. Denoting fields on $\mathcal C_+$
by $\varphi_+(x)$ and on $\mathcal C_-$ by $\varphi_-(x)$, the
initial fields are fixed by $\varphi_1(\bx) = \varphi_+(t=0,\bx)$
and $\varphi_2(\bx) = \varphi_-(t=0,\bx)$. Non-equilibrium
correlation functions are obtained by functional differentiation
and setting $J=0$.

With this notation the expectation value of a real-time observable
${\cal A}(\varphi)$ can be written as
\begin{eqnarray}
\avr{{\cal A}(\varphi)} &=& \int {\rm d} \varphi_1 {\rm d}
\varphi_2\, \rho(\varphi_1,\varphi_2) \int_{\varphi_+(0)=\varphi_1}^{\varphi_-(0)=\varphi_2}
D^{\prime}\varphi_- D^{\prime}\varphi_+
e^{iS[\varphi_+]-iS[\varphi_-]}
{\cal A}(\varphi_+)\\
&=&\int D\varphi_- D\varphi_+ e^{iS_\rho[\varphi_+,\varphi_-]}
{\cal A}(\varphi_+)\,.
\end{eqnarray}
Here $S_\rho$ contains the actions on both contour branches as
well as the density operator. In the following we consider
Gaussian initial density matrices for which, neglecting all space
dependence for simplicity, the most general $S_\rho$ reads~\cite{Berges:2004yj}
\begin{eqnarray}
S_\rho[\varphi_+,\varphi_-]&=&S[\varphi_+]-S[\varphi_-]
-\frac{i}{a_t}S_0(\varphi_+(t=0),\varphi_-(t=0)]\label{eq:Sp}\,,\\
S_0[\varphi_+,\varphi_-]&=& i\dot\phi(\varphi_+-\varphi_-)
-\frac{\sigma^2+1}{8\xi^2}\left((\varphi_+-\phi)^2+(\varphi_--\phi)^2\right)\nonumber\\
&+&\frac{i\eta}{2\xi}\left((\varphi_+-\phi)^2-(\varphi_--\phi)^2\right)\nonumber\\
&+&\frac{\sigma^2-1}{4\xi^2}(\varphi_+-\phi)(\varphi_--\phi) \, .
\end{eqnarray}
The real parameters $\phi$, $\dot\phi$, $\sigma$, $\xi$, $\eta$ determine a
complete set of independent initial one-point and two-point
correlation functions:
\begin{eqnarray}
\phi&=&\avr{\varphi(t=0)} \quad , \qquad \dot\phi \, = \,
\avr{\dot\varphi(t=0)} \, ,\nn
\xi^2&=&\avrc{\varphi(t=0)\varphi(t=0)}\, ,\nn
\eta\xi&=&\frac12\avrc{\dot\varphi(t=0)\varphi(t=0)+\varphi(t=0)\dot\varphi(t=0)}\,
, \nn
\eta^2+\frac{\sigma^2}{4\xi^2}&=&\avrc{\dot\varphi(t=0)\dot\varphi(t=0)}
\, . \label{eq:initialcond}
\end{eqnarray}

Starting from a given initial density matrix the nonequilibrium
simulation is carried out using the Langevin equation
(\ref{eq:complexlange}) with the action replaced by (\ref{eq:Sp}),
i.e.
\begin{equation}
\varphi_\pm'(x) = \varphi_\pm(x) + i\, \epsilon \, \frac{\delta
S_\rho[\varphi_+,\varphi_-]}{\delta \varphi_\pm(x)} +
\sqrt{\epsilon}\, \eta_\pm(x) \, , \label{eq:nonequilange}
\end{equation}
updating all the points including $\varphi_\pm(t=0)$.

\begin{figure}[t]
\begin{center}
\epsfig{file=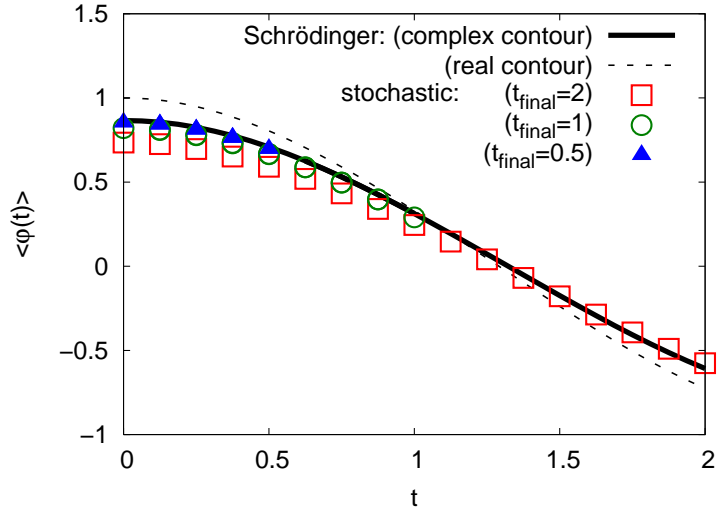,width=10.cm}
\end{center}
\caption{Non-equilibrium time evolution of the field expectation
value as a function of time. Shown are results for different
real-time extents of the lattice. Shorter $t_{\rm final}$ lead to
improved agreement with results of the Schr{\"o}dinger equation on
the employed complex contour.} \label{fig:tx1}
\end{figure}

As an example, Fig.~\ref{fig:tx1} shows the time evolution of the
expectation value $\avr{\varphi(t)}$ as a function of time, all in
units of the mass parameter $m$ of equation (\ref{eq:effpot}).
Here the coupling is $\lambda = 1$ for zero spatial dimension. The
initial conditions are $\avr{\varphi(t=0)}=1$,
$\avrc{\varphi(t=0)\varphi(t=0)}=1$,
$\avrc{\dot\varphi(t=0)\dot\varphi(t=0)}=0.25$ and zero for the
remaining quantities of (\ref{eq:initialcond}). The time contour
is tilted with $5\%$ slope, which is denoted as complex contour in
Fig.~\ref{fig:tx1}. Results using stochastic quantization are
given for three different lengths of the real-time extent of the
contour ($t_{\rm final}=2,1$ and $0.5$). For comparison we also
show the results of the numerical solution of the corresponding
Schr{\"o}dinger equation (see Sec.~\ref{sec:thermal}) for real
times as well as for the complex times on the tilted contour. This
also illustrates the contour dependence of the results. 
One observes that the stochastic quantization
algorithm produces accurate results on the contour for
sufficiently short real-time extent. Like for the thermal
equilibrium results of Sec.~\ref{sec:thermal}, we find that it
becomes inaccurate for late-time evolution.

\section{Accuracy tests II: nonabelian gauge theory}
\label{sec:testgauge}

In the following we perform a similar investigation for $SU(2)$
gauge theory as done above for the scalar theory. Though some
aspects of the application of real-time stochastic quantization
are comparable, we will find that there are crucial additional
restrictions concerning its range of validity for nonabelian gauge
theories.

The equation for the Langevin evolution is given by
Eq.~(\ref{eq:ALangevinM}) for the action (\ref{eq:clgaugeaction}),
where we employ the approximation that $g_0 = g_s$. We use the $
U=a{\bf 1} + i b_a\sigma^a,\ a^2+b^2_a=1$ representation for
$SU(2)$ matrices. For $SL(2)$, the representation with the same
constraint can be used, but the parameters $a$ and $b_a$ are no
longer real. We expand the exponential to first order in
$\epsilon$, which means we must include the square of the noise
term, which is proportional to unity. The evolution equation then
reads
\begin{eqnarray}
U^{\prime}_{x, \mu} &=& \left( a{\bf 1}   + i \sigma_a
\left(\epsilon\, i D_{x \mu a} S[U] + \sqrt{\epsilon}\, \eta_{x
\mu a} \right) \right) U_{x, \mu}\, . \label{eq:Alattice}
\end{eqnarray}
To stay in group space the constant $a$ is calculated from the
constraint $ a^2 + b_a^2=1$ of the matrix transforming $U_{x,
\mu}$.

As the starting configuration for the Langevin-time evolution we
take all link variables equal to unity. Typically we employ $ 1/
\gamma= a_t/a_s = 0.01-0.5 $. We use lattices of spatial size
$N^3=4^3$. The Langevin step used is $\sim 10^{-6}$. All
quantities are given in units of $a_s$. The triangle contour of
Fig.~\ref{fig:contourev} is used, with $\tau_+$ being the
imaginary extent of the contour on the forward part and $\tau_-$
on the backward part. We calculate thermal distributions, with
inverse temperature $\tau_+ + \tau_-=4$.
\begin{figure}[t]
\begin{center}
\epsfig{file=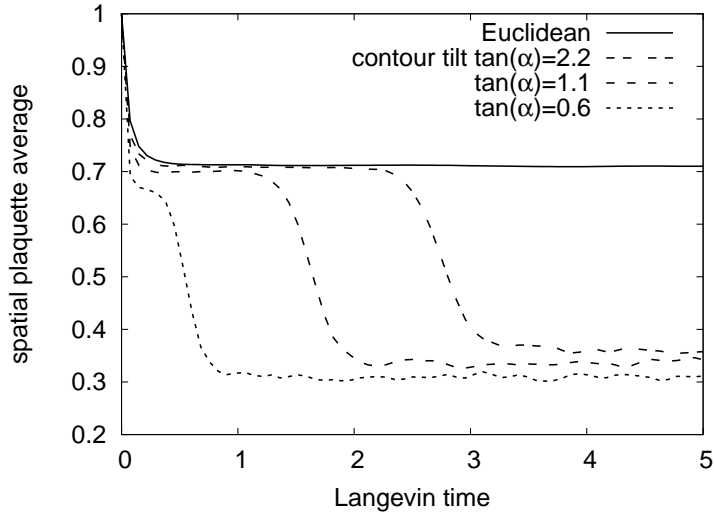,width=10.cm}
\end{center}
\caption{The spatial plaquette average as a function of Langevin
time. Shown are results for different complex contours. Here
$\alpha = 0$ corresponds to a contour with an infinite extent
along the real-time axis, while $\alpha = \pi/2$ denotes the
Euclidean contour. The longer the real-time component of the
contour the less accurately the thermal solution is reproduced.}
\label{fig:2fixp}
\end{figure}

Fig.~\ref{fig:2fixp} shows the {\em Langevin-time} evolution of
the spatial plaquette average.\footnote{The plaquettes are also
averaged in real (complex) time.} The solid line shows the result
for vanishing real-time extent, i.e.\ for a Euclidean contour. The
different dashed curves correspond to results for complex contours
on isosceles triangles ($\tau_+=\tau_-=2$) each having a different
tilt $\alpha$ with respect to the real-time axis. Here $\alpha =
0$ would correspond to an infinite extent along the real axis. One
observes that with increased real-time extent or smaller tilt
$\alpha$ the correct thermal solution is approached less
accurately.

In particular, it is only approached at intermediate
Langevin-times, irrespective of the details of the non-vanishing
real-part of the contour. This aspect differs from the scalar case
where short real-time extents lead to stable thermal solutions.
For the nonabelian gauge theory the correct thermal fixed point is
approached at first. However, it is not stable and the Langevin
flow exhibits a crossover to another (stable) fixed point. The
fact that the thermal solution is not a true fixed point for
non-Euclidean contours can be seen by monitoring for example
$\left\langle\textrm{Im}(i\textrm{Tr}(U_{x,\mu}
\sigma^a))^2\right\rangle$, which is zero for $SU(2)$ matrices in
the Euclidean theory. It is found to grow exponentially while
observables of the original $SU(2)$ theory such as the plaquette
average still approach the thermal solution.
\begin{figure}[t]
\begin{center}
\epsfig{file=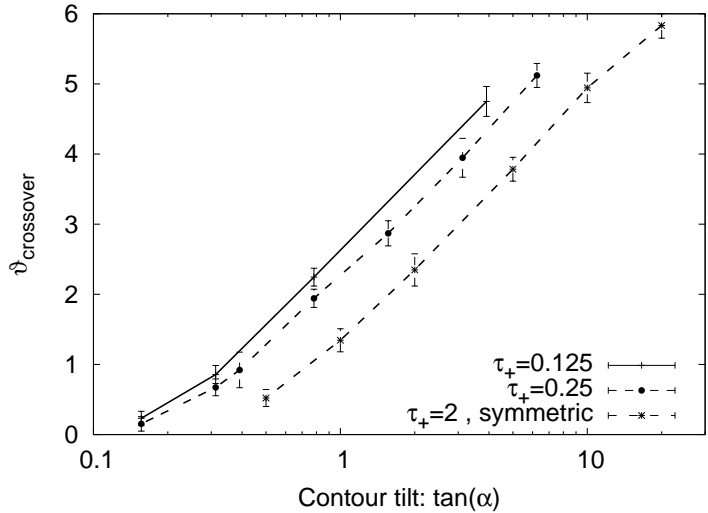,width=10.cm}
\end{center}
\caption{The Langevin-time $\vartheta_{\rm crossover}$ for the
crossover away from the approximate thermal solution to another
fixed point. (See Fig.~\ref{fig:2fixp}.) It is displayed as a
function of ${\rm tan}\alpha$ for various time contours. Contours
with different real-time extent, but with the same $\tau+$ are
connected with a line.} \label{fig:cross}
\end{figure}

This is quantified in Fig.~\ref{fig:cross}, where the
Langevin-time $\vartheta_{\rm crossover}$, at about which the
crossover from the approximate thermal solution to another fixed
point occurs, is given for various time contours. One observes
that the time of the crossover is mostly dependent on the angle of
the slope of the contour. Connecting points with the same
$\tau_+$, one sees that $ \theta_{\rm crossover} $ is
approximately proportional to the logarithm of the tangent of the
tilt of the contour. Here $\vartheta_{\rm crossover}$ was measured
on $ N_t=16 $ lattices, using $\approx 5$ runs per parameters.
\begin{figure}[t]
\begin{center}
\epsfig{file=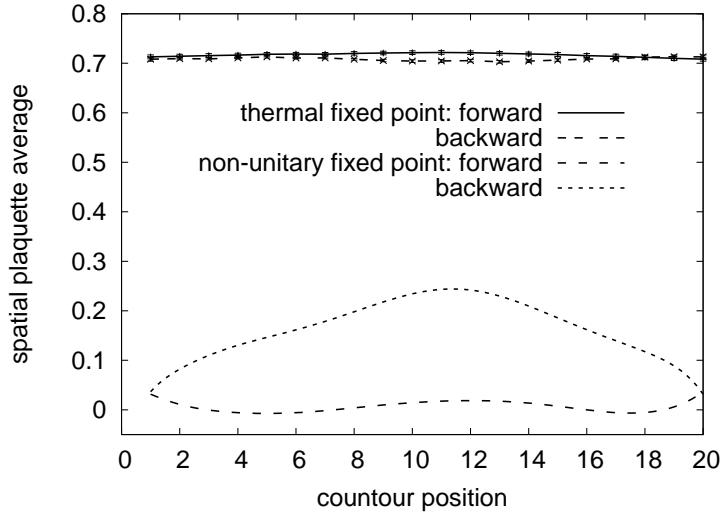,width=10.cm}
\end{center}
\caption{The spatial plaquette averages as a function of the index
of lattice sites along the time-contour. One observes that it is
time translation invariant to rather good accuracy for the
approximate thermal fixed point, while it is not for the
non-unitary fixed point.  } \label{fig:realtime}
\end{figure}

In Fig.~\ref{fig:realtime} the spatial plaquette average is shown
as a function of the index of lattice sites along the
time-contour. While the Langevin-time evolution approaches the
thermal solution it is time translation invariant to rather good
accuracy. The observed small discrepancy is decreasing with
Langevin-time while the system stays close to the approximate
thermal solution. In contrast, for the stable fixed point at late
Langevin-time the plaquette averages are not time translation
invariant along the contour --- at least for the finite
Langevin-times for which we followed the evolution. Here we
employed a contour with $\tau_+=0.5$ and $\tau_-=3.5$ with
$a_t/a_s=0.03$ and $N_t=20$ from an average over 10 runs. The
real-time evolution near the thermal solution has been averaged over
Langevin time $0.75<\vartheta<1.25$, for the non-unitary fixed point
it has been averaged over $3.75< \vartheta < 4.25 $. 
\begin{figure}[t]
\begin{center}
\epsfig{file=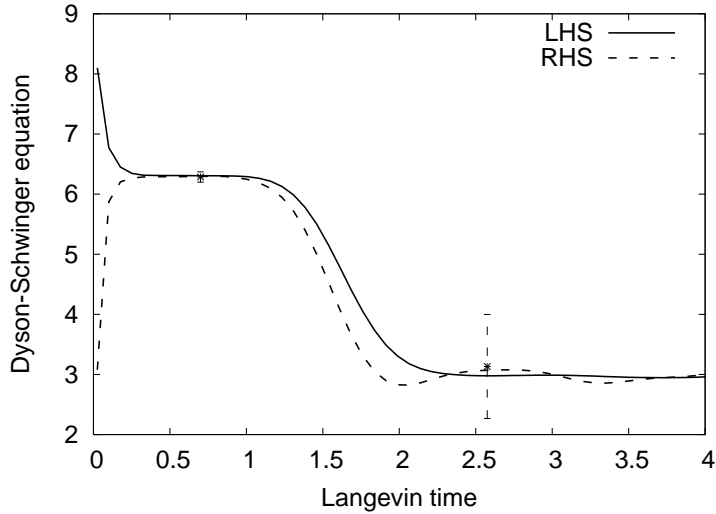,width=10.cm}
\end{center}
\caption{Numerical check of the Dyson-Schwinger equation for a
spatial plaquette variable. Displayed are separately the LHS and
the RHS of (\ref{eq:DSpre}).} \label{fig:DSnum}
\end{figure}

As is shown is Sec.~\ref{sec:DSU}, all converging solutions of the
stochastic dynamics method fulfil the same infinite set of
(symmetrized) Dyson-Schwinger identities of the quantum field
theory. This is remarkable in view of the different ``physical''
and ``unphysical'' solutions that are observed. In
Fig.~\ref{fig:DSnum} this is visualized for the example of the
Dyson-Schwinger equation for a spatial plaquette variable. Plotted
are separately the LHS and the RHS of the Dyson-Schwinger equation
(\ref{eq:DSpre}) for the plaquette (see also Fig.~\ref{fig:DS}).
The plot displays the respective LHS and RHS as a function of
Langevin-time. The flow with Langevin-time quickly leads to a
rather accurate agreement of both sides such that the
Dyson-Schwinger equation is fulfilled. However, after some
Langevin-time they start deviating again, finally leading to
another stationary value where the LHS and RHS agree to reasonable
accuracy. Here the fluctuations of measured quantities of the
system are much bigger. This is indicated by the given typical
statistical error bars, with a comparably huge statistical error
for the late Langevin-time evolution. We used median averaging for
the right hand side of the equations. For the contour we employ
$\tau_+=\tau_-=2$ with $N_t=8$ and $a_t/a_s=0.25$.

\section{Conclusions}
\label{sec:conclusions}

The motivation for this paper is the question of making real-time,
nonequilibrium quantum field theory amenable to numerical simulations 
from first principles. This would not only boost our knowledge directly, 
it would also allow better testing of approximate analytical tools.
This concerns, in particular, strongly interacting theories such as QCD,
where reliable analytical approximations are difficult to find at 
phenomenologically relevant energies. Despite its importance research 
on real-time lattice gauge theory is still in its infancies. 
It has been a delicate problem and the various attempts based on
reweighting or on Euclidean simulations have encountered major 
difficulties. Our aim here was to study the applicability of stochastic 
quantization to real-time, nonequilibrium problems. This method does, 
a priori, not involve any reweighting, nor redefinition of the 
Minkowski dynamics in terms of an associated Euclidean one.

For setting up the procedure, both for scalar and for
nonabelian gauge theory, we started from a five-dimensional classical
Hamiltonian dynamics supplemented by a stochastic description and 
correctly accounting for the symmetries of the models. We then showed
that the fixed points of the Langevin (fifth-time) dynamics 
fulfil the infinite set of Dyson-Schwinger equations 
associated with the respective quantum theory. 
We established in this way a set of identities to be 
fulfilled by the expectation values. These general results 
both allow checks of the simulations and suggest 
means to understand convergence and metastability problems.

In the second part we undertook a numerical study of the real-time stochastic 
quantization approach applied to scalar and $SU(2)$ gauge theory 
as a paradigmatic Yang Mills model. The main aim being to
understand the problems of, and to develop means to control the method.
Beyond varying the parameters of the implementation we also
worked on various realizations of nonequilibrium and of non-zero temperature 
problems, which correspondingly define various integration contours. 
Our findings can be summarized as follows (both for scalar and
Yang Mills fields, unless explicitely distinguished):
\par \medskip
\noindent - Instabilities of the Langevin dynamics 
are controllable: if the Langevin step size is chosen small enough 
run-away trajectories are rather seldom and the results do not suffer
from discarding them.\par \medskip
\noindent - Tilting the integration contour in the complex plane is 
a gauge invariant way to improve both the
convergence and the accuracy of the results. The physical effect of
this tilt can be understood from exact results (e.g., Schr{\"o}dinger 
equation for the anharmonic oscillator) and small tilts can therefore
be used in the simulation.\par \medskip
\noindent - Short real-time physics in thermal equilibrium can be 
reproduced reasonably well
if the length of the real time contour is small on
the scale of the 
inverse temperature $\beta$, the details depending also on other 
parameters as explained in the main text. 
The Langevin flow  is here dominated by a thermal fixed point which 
fulfils the unsymmetrized Dyson-Schwinger identities and 
describes a physical solution.
This fixed point is stable for 
the scalar theory while it is metastable for the gauge theory, 
its life-time depending then on the
contour and the other parameters of the problem. \par \medskip
\noindent - For longer contours the boundary
conditions in physical time do not seem to constrain enough the Langevin flow
and the life-time of the thermal (physical) fixed point decreases 
(for gauge theory),
or the fixed point becomes fully unstable (for scalar theory).
A second, apparently stable fixed point develops for large
Langevin times. This latter fixed point represents an
unphysical, non-unitary regime, to be recognized by non-translational 
invariant expectation functions and violation of 
the unsymmetrized Dyson-Schwinger identities (while the symmetrized ones
are still satisfied, indicating convergence).\par \medskip
\noindent -  For nonequilibrium problems similar observations 
hold. For the scalar model all these findings have been checked by
comparing with results from a numerical solution of the corresponding 
Schr{\"o}dinger equation.\par\medskip
In view of these results we conclude that the Langevin approach 
can be used in direct numerical simulations of Minkowski 
theories, including Yang Mills theory, if a number
of (rather severe) restrictions are observed.
In those situations where physical solutions become
unstable other, unphysical solutions will develop, but tests for 
differentiating them can typically be devised and applied.
We did not try to optimize so far. The method allows for 
quite some flexibility as to which quantities
are chosen to define the stochastic process, or introducing
a stochastic reweighting. This is work in progress.  
\par\medskip
Part of the numerical calculations have been performed on HELICS of the IWR Heidelberg.

\end{document}